\documentclass[12pt]{article}

\usepackage{scicite}
\usepackage{graphicx}
\usepackage{color}
\usepackage{soul}
\usepackage{url}
\usepackage{multirow}
\usepackage{setspace}
\usepackage{changepage}
\usepackage{ragged2e}
\usepackage{anyfontsize}

\usepackage{times}

\newcommand{\heii}{He\,\textsc{ii}}
\newcommand{\civ}{C\,\textsc{iv}}

\def\apj{Astrophys. J.}

\def\mnras{Mon. Not. R. Astron. Soc.}
\def\pasj{Pub. Astro. Soc. J.}
\def\apjs{Astrophys. J. Suppl. Ser.}
\def\araa{Annu. Rev. Astron. Astrophys.}
\def\aap{Astron. Astrophys.}
\def\nat{Nature}
\def\xgtia{X. G. Tel. Ins. Astron.}
\def\spiecs{S. Pho. Ins. Eng. Con. Ser.}

\newenvironment{sciabstract}{%
\begin{quote} \bf}
{\end{quote}}

\title{Inspiraling streams of enriched gas observed around a massive galaxy 11 billion years ago}
\author
{
\renewcommand*{\thefootnote}{\fnsymbol{footnote}}
\setcounter{footnote}{1}
{\fontsize{0.45cm}{0.4cm}\selectfont Shiwu Zhang,$^{1}$\footnotemark \ \setcounter{footnote}{1} Zheng Cai,$^{1,2\ast}$\footnote{These authors contributed equally to this work} \ Dandan Xu,$^{1}$ Rhythm Shimakawa,$^{3,4}$}\\
{\fontsize{0.45cm}{0.4cm}\selectfont Fabrizio Arrigoni Battaia,$^{5}$
Jason Xavier Prochaska,$^{6,7}$  Renyue Cen,$^{8,9}$} \\ 
{\fontsize{0.45cm}{0.4cm}\selectfont Zheng Zheng,$^{10}$ Yunjing Wu,$^{1}$ Qiong Li,$^{11,12}$ Liming Dou,$^{13}$ Jianfeng Wu,$^{14}$}\\  
{\fontsize{0.45cm}{0.4cm}\selectfont Ann Zabludoff,$^{15}$ Xiaohui Fan,$^{15}$ Yanli Ai,$^{16}$  Emmet Gabriel Golden-Marx,$^{1}$ Miao Li,$^{17}$}\\
{\fontsize{0.45cm}{0.4cm}\selectfont Youjun Lu,$^{18,19}$ Xiangcheng Ma,$^{20}$ Sen Wang,$^{1}$ Ran Wang,$^{11}$ Feng Yuan$^{21}$}\\
\small{$^{1}$Department of Astronomy, Tsinghua University, Beijing 100084, China}\\
\small{$^{2}$School of Mathematics and Phyisics, Qinghai University, Xining 810016, China}\\
\small{$^{3}$National Astronomical Observatory of Japan, National Institutes of Natural Sciences, Tokyo 181-8588, Japan}\\
\small{$^{4}$Waseda Institute for Advanced Study, Waseda University, Tokyo 1690051, Japan}\\
\small{$^{5}$Max-Planck-Institut f\"{u}r Astrophysik, Garching bei M\"{u}nchen D-85748, Germany}\\
\small{$^{6}$Department of Astronomy and Astrophysics, University of California, Santa Cruz, CA 95064, USA}\\
\small{$^{7}$Kavli Institute for the Physics and Mathematics of the Universe, The University of Tokyo, Kashiwa, 277-8583, Japan}\\
\small{$^{8}$Department of Physics, Zhejiang University, Hangzhou 310027, China}\\
\small{$^{9}$Department of Astrophysical Sciences, Princeton University, Princeton NJ08544, USA}\\
\small{$^{10}$Department of Physics and Astronomy, University of Utah, Salt Lake City UT84112, USA}\\
\small{$^{11}$Kavli Institute for Astronomy and Astrophysics, Peking University, Beijing 100871, China}\\
\small{$^{12}$Jodrell Bank Centre for Astrophysics, University of Manchester, Manchester M13 9PL, UK}\\
\small{$^{13}$Department of Astronomy, Guangzhou University, Guangzhou 510006, China}\\
\small{$^{14}$Department of Astronomy, Xiamen University, Xiamen 361005, China}\\
\small{$^{15}$Steward Observatory, University of Arizona, Tucson AZ85721, USA}\\
\small{$^{16}$College of Engineering Physics, Shenzhen Technology University, Shenzhen 518118, China}\\
\small{$^{17}$Center for Computational Astrophysics, Flatiron Institute, New York NY10010, USA}\\
\small{$^{18}$National Astronomical Observatories, Chinese Academy of Sciences, Beijing 100101, China}\\
\small{$^{19}$School of Astronomy and Space Sciences, University of Chinese Academy of Sciences, Beijing 100049, China}\\
\small{$^{20}$Department of Astronomy and Theoretical Astrophysics Center,}\\
\small{University of California Berkeley, Berkeley CA94720, USA}\\
\small{$^{21}$Shanghai Astronomical Observatory, Chinese Academy of Sciences, Shanghai 200030, China}\\
\\
\small{$^\ast$Correspondence to: zcai@mail.tsinghua.edu.cn}
}

\date{}

\begin{document} 

\baselineskip24pt


\maketitle


\begin{sciabstract}
{Stars form in galaxies from gas that has been accreted from the intergalactic medium.}
Simulations have shown that recycling of gas—the reaccretion of gas that was previously ejected from a galaxy—could sustain star formation in the early Universe.
We observe the gas surrounding a massive galaxy at redshift 2.3 and detect emission lines from neutral hydrogen, helium and ionized carbon that extend 100 kiloparsecs from the galaxy. 
The kinematics of this circumgalactic gas is consistent with an inspiraling stream. The carbon abundance indicates that the gas had already been enriched with elements heavier than helium, previously ejected from a galaxy. We interpret the results as evidence of gas recycling during high-redshift galaxy assembly.

\end{sciabstract}


Simulations of galaxy formation in the early Universe indicate that low-mass galaxies grow by the direct accretion of gas from the circumgalactic medium (CGM) and intergalactic medium (IGM) \cite{Tumlinson2017}. 
Both simulations and observations show that galaxies in low mass dark matter halos with a halo mass of $M_{\rm h}<10^{12} \ M_{\odot}$ \cite{Keres2005,Stern2020} can accrete streams of gas at $10^{4}$ K,
linking to the surrounding CGM and IGM by a web of pristine gas (gas with almost no metals) filaments (where $M_{\odot}$ denotes the solar mass) \cite{Keres2005,Tumlinson2017}. 
Transport of gas along streams prevents it from being shock-heated while falling into the potential well of the dark matter halo, so this process is referred to as `cold mode' accretion \cite{Keres2005}. 
Cold-mode accretion can explain the high star formation rate (SFR) of high-redshift galaxies, 
 and the angular momentum growth of galaxies halos\cite{Stewart2017}.

Although cold-mode accretion is expected for pristine gas, predictions differ for metal-enriched gas (gas with higher abundances of elements heavier than helium, referred to as its metallicity).
Cosmological simulations predict potentially-observable quantities of
metal-enriched CGM around galaxies with  $M_{\rm h}10^{12} M_{\odot}$ \cite{Suresh2019,Lehner2019}. 
Because metal-enriched CGM gas can cool more efficiently than pristine gas, metal-enriched accretion (recycled inflow) could provide additional gas and boost the SFR of galaxies in massive halos ($M_{\rm h}>10^{12} \ M_{\odot}$) at $z>2$ \cite{Oppenheimer2010,Angles-Alcazar2017,Brennan2018,Grand2019}.
Observations of absorption lines towards background sources have implied the presence metal-enriched CGM around galaxies \cite{Prochaska2014},  but these only provide information at a single point. 
To determine the spatial distribution of CGM gas requires studying its emission lines.

{\bf \large Observations of MAMMOTH-1} 

Lyman alpha (Ly$\alpha$) is an emission line of neutral hydrogen that has a rest-frame wavelength of 1216 \AA.
Observations of enormous Ly$\alpha$ nebulae with wide-field integral field spectrographs,  could determine the physical properties and kinematics of the CGM at $z>2$. 
One such Ly$\alpha$ nebula is the MAMMOTH-1 (J2000 $14^{\rm h}41^{\rm m}24.42^{\rm s}$, $+40^{\circ}03'09.7''$) nebula at $z\approx2.31$, which has Ly$\alpha$ emission with a projected spatial extent of 442 kpc and Ly$\alpha$ luminosity of $L_{\rm Ly\alpha}=5.1 \pm 0.1 \times 10^{44}$ erg s$^{-1}$ \cite{Cai2017a}. 
It resides in an overdense galaxy environment \cite{Cai2017a, Arrigoni2018a, Emonts2019}.  
Our team observed MAMMOTH-1 with the Keck Cosmic Web Imager (KCWI) on the 10-m Keck II telescope, in imaging 
spectroscopy mode centered on the Ly$\alpha$, \civ \ $\rm 1548/1550 \ \AA$, and \heii \ $\rm 1640 \ \AA$ emission lines. 
 We also performed narrowband imaging of redshifted H$\alpha$ (line of neutral hydrogen with the rest-frame wavelength of 6563 \AA) emission using the Multi-Object InfraRed Camera and Spectrograph (MOIRCS) on the 8-m Subaru telescope. 
 We supplement these data with archival observations at near infrared, x-ray, and radio wavelengths \cite{Methods}.

 We optimally extracted images from the KCWI data (Fig.~1, A to C) \cite{Arrigoni2018a, Methods}. 
 These show that the flux peaks of the emission lines Ly$\alpha$, \civ,\ and \heii \ coincide with a quasar detected in the x-ray data (fig.~S1B), which we designate G-2 (J2000 $14^{\rm h}41^{\rm m}24.42^{\rm s}$, $+40^{\circ}03'09.7''$)\cite{Methods}.
 G-2 is located inside the nebula and provides the ionising photons that excite the gas emission lines.
The Ly$\alpha$, \civ, and \heii \ emission regions are asymmetrically distributed around G-2. 
 Both \heii\ and \civ\ are spatially extended, with projected scales of 88 and 108 kpc (2-$\sigma$ emission), respectively, and have luminosities of {\color{black} $L_{\rm HeII}=5.48\pm 0.13 \times 10^{42}$ erg s$^{-1}$} and {\color{black} $L_{\rm CIV}=10.50\pm 0.16\times 10^{42}$ erg s$^{-1}$}, respectively. 
The MOIRCS narrowband imaging \cite{Methods} shows that the H$\alpha$ emission has a projected scale of 97 
kpc, 
with a luminosity of $L_{\rm H\alpha}=2.33\pm 0.15 \times 10^{43}$ erg s$^{-1}$. 
 From the archival CO (J=1→0) radio observations \cite{Emonts2019},
 we measure the redshift of G-2 to be $z=2.3116\pm0.0004$ \cite{Methods}.
The CO (J=1→0) and CO (J=3→2) observations (fig.~S1A) \cite{Emonts2019,Li2021} show that the sources around G-2 marked as G-1, G-3, G-4, G-5, and G-6 (Fig.~1) also have redshifts of $z\approx 2.31$ and thus are located within the nebula.


 We used the spectral information to construct a flux-weighted Ly$\alpha$ velocity map. 
 This shows two regions of gas, each with gas at a similar velocity extending for $\geq100$ kpc, which we refer to as regions A and B.
The ionizing source, G-2, is located within region A.
One-dimensional (1D) spectra (Fig.~2) extracted from selected apertures (labeled in Fig.~1E) have double- or triple-peaked structures, which we fitted with models consisting of multiple Gaussians. 
The Ly$\alpha$ and the \heii \ emission have similar double- or triple-peaked at each location, indicating that the line profiles are attributable to the motion of the ionized gas rather than Ly$\alpha$ radiative transfer effects \cite{Yang2014}.
Using smaller apertures (fig.~S2) confirms this result \cite{Methods}.
In the large-scale CGM where no \heii\ is detected, 
 we expect radiative transfer effects to be weaker, 
so the Ly$\alpha$ also traces the cool gas kinematics in those regions \cite{Methods}. 

{\bf \large Analysis of line ratios} 

 Using the diffuse \civ, \ \heii, \ and H$\alpha$ emission lines, we studied the properties of the CGM with spatially resolved line ratio diagnostics \cite{Methods}. 
H$\alpha$ emission is better suited for line ratio diagnostics than Ly$\alpha$ because it is less affected by 
resonant scattering \cite{Arrigoni2015a}.
H$\alpha$ is also less effected by dust attenuation; we find that any effect on the line ratios is negligible \cite{Methods}.
We consider two possible emission mechanisms.
In the pure photoionization scenario, the gas is highly ionized by strong ultraviolet radiation from the quasar, causing the emission lines to be dominated by recombination. 
In the alternative shock-with-precursor scenario, the gas moves at a velocity higher than the local sound speed, so a shock front forms at the leading edges of the clouds. 
Such shocks could heat the gas to a sufficient temperature that it cools by emitting soft x-ray radiation \cite{Allen2008}. 
Model line ratios are shown in fig.~S5 for the pure photoionization scenario and in Fig.~3 for the shock-with-precursor scenario. 
The observed \civ/H$\alpha$ ratio is consistent with both scenarios, whereas the \heii/H$\alpha$ ratio can only be produced by the shock-with-precursor scenario  — it is an order of magnitude smaller than predicted for the pure photoionization model. 
We examined alternative assumptions about the quasar emission and dust attenuation, finding that a photoionization scenario is unlikely to produce the observed line ratios in all cases \cite{Methods}.
We therefore conclude that the shock-with-precursor scenario is the mechanism responsible for the observed emission lines.

From the line ratio diagnostics, we find that the CGM metallicities are as high as solar metallicity ($Z_\odot$) within 2-$\sigma$ range (Fig.~3). 
This is an order of magnitude higher than the previously measurements of the metallicity of the interstellar medium in other $M_{\star}\le 10^9\ M_\odot$ galaxies 
at $z\approx$2 to 3 \cite{Steidel2014,Tumlinson2011} but consistent with the CGM metallicity of some galaxies at $z\approx0$ within 2-$\sigma$ range \cite{Wotta2019}. 
Previous studies have found that metal-enriched gas, traced by \civ \ absorption toward background source, is distributed within $\approx125$ kpc around galaxies at $z=$2 to 3 \cite{Steidel2010}. 
Other studies have shown that CGM metallicity could be higher than $Z_{\odot}$ at $z=1$ \cite{Lehner2016}.
Our measurements show that the CGM in MAMMOTH-1 has been enriched to between 0.1 and 1.0$Z_\odot$ on $\approx 100$ kpc scales.
{\bf \large Interpretation of the kinematics} 

We investigated whether the observed kinematics could be produced by an active galactic nucleus (AGN) outflow (see supplementary text in the supplementary materials). 
We find that they cannot, for two reasons. 
First, we expect outflow to decelerate as it propagates in CGM, as a result of energy loss \cite{Fabian2012,Richings2018,Circosta2018}. 
Instead, the observed line-of-sight velocity profile (Fig.~4F) shows that the redshifted velocity is roughly constant and that the absolute value of the blueshifted velocity increases with increasing distance from G-2.
Second, if the kinematics were due to an AGN outflow, the observations would require an implausibly high coupling efficiency between the outflow power and the AGN luminosity (supplementary text) that is almost one order of magnitude greater than found by previous observations or simulations (fig.~S13). 

Alternatively, the gas metallicity, spatial distribution and kinematics could be due to metal-enriched inspiraling streams. 
Gas in the CGM could have been enriched and expelled by previous outflows, 
from the AGN or starbursts in other galaxies.
Cosmological simulations predict that inspiraling cool streams allow the accreting galaxies to gain high angular momentum \cite{Stewart2011, Stewart2017}.
In cosmological simulations \cite{Nelson2019},  $\sim$46\% of halos with masses $M_{\rm h}\geq 10^{13} \ M_{\odot}$ at $z=2$ have inspiraling streams that are metal-enriched on CGM scales \cite{Methods}.
Those simulations predict that, in addition to pristine gas accretion, accretion of recycled, metal-enriched gas could be a common process.

Motivated by those simulations, 
we constructed a simple kinematic model of metal-enriched inspiraling streams to interpret our observations \cite{Methods}. 
The model consists of three inspiraling streams: two that surround G-2 and a third that surrounds G-5. 
The geometry of the model is shown in Fig.~4C. 
Comparing this model to the data, we find that both the simulated velocity map (Fig.~4B) and line profiles (Fig.~4, D and E) are consistent with the observations. 
The reduced $\chi^{2}$ between the simulated and observed spectra is  $\sim$0.9. 
The kinematics is also reproduced by the model (Fig.~4F), within the 1-$\sigma$ scatter.
{\bf \large Implications for gas accretion}

Cosmological simulations have also shown that recycled inflows can provide gas accretion at $z\sim0$ \cite{Oppenheimer2010,Grand2019}. 
Other simulations \cite{Angles-Alcazar2017} have shown that, at $z=2$, the fraction of stellar mass contributed by recycled gas is 40\% in a $10^{12.8} M_{\odot}$ halo.
Our interpretation of the MAMMOTH-1 observations are consistent with the latter scenario \cite{Angles-Alcazar2017}.

We calculate that the streams provide a mass inflow rate $\dot{M}_{\rm in}=703^{+101}_{-78} \ M_{\odot}$ year$^{-1}$ \cite{Methods}, which is higher than the SFR of G-2 ($81\pm18 M_{\odot}$ year$^{-1}$) derived from its far-infrared emission \cite{Methods}. 
We suggest that this indicates link between the accretion rate of the recycled gas inflow and the SFR of G-2 \cite{Methods}.
The observed line-of-sight velocity profile (Fig. 4F) is consistent with cool gas undergoing 
deceleration as it falls into the dark matter halo, as predicted by semianalytic models of cool CGM gas \cite{Lan2019,Afruni2019}.
In those semianalytic models, the deceleration is interpreted as a consequence of drag forces exerted by the hot coronal gas on the cool gas stream. 
Because MAMMOTH-1 resides in the peak of overdensity that is also a galaxy group, the diffuse metal-enriched gas could also arise from galaxy interactions, such as tidal stripping. 
We estimate that this scenario is negligible 
on the CGM scale (supplementary text). 
A Hubble Space Telescope image of MAMMOTH-1 shows no evidence of tidal-stripping on large scales (Fig.~1K).

Our spectroscopic observations show that the group of nearby galaxies has a redshift gradient qualitatively consistent with that of the CGM (Fig.~1, D and table~S3). 
This indicates that the large-scale orbital angular momentum of the galaxy group aligns with the CGM angular momentum. 
Any satellite galaxies moving around G-2 could impart angular momentum to the enriched cool CGM gas, which would then flow back to the galaxy in an inspiraling stream (fig.~S8A).
The CGM gas flow would then induce a shock, facilitating gas cooling through line emission. 
From the estimated inflow rate of the recycled gas, the recycled inflow could sustain star formation in G-2 at this redshift. 

\newpage
{\bf \Large Acknowledgments} 

We thank the anonymous referees for insightful comments that substantially improved the manuscript. 
Z.C. and S.Z. thank S. Cantalupo, H.-W. Chen, M. Gronke, K. Kakiichi, C. Martin, M. Matuszewski, V. Springel, Y. Su and H. Zhou for useful discussions.
Z.C. and S.Z. also thank J. Zou and B. Wang for proofreading the manuscript.

{\bf Funding:} Z.C., S.Z., and Y.W. are supported by the National Key R\& D Program of China (grant No. 2018YFA0404503), the National Science Foundation of China (grant 12073014). 
D.X. and S.W.  are supported by the Tsinghua University Initiative Scientific Research Program (grant 2019Z07L02017). 
R.S. acknowledges support from Grants-in-Aid for Scientific Research (KAKENHI; 19K14766) through the Japan Society for the Promotion of Science (JSPS).
Z.Z. is supported by NSF grant AST-2007499.  A.Z. acknowledges funding from NSF grant AAG-1715609.

{\bf Author contributions:}
Z.C. conceived the project. S.Z. and Z.C. led the data reduction, analysis, and manuscript writing of the manuscript. 
 Z.C., R.S., J.X.P., X.F. and Q.L. led the telescope proposals, with Z.C. as the principal investigator of most of the proposals. 
 R.S. contributed to the observations and data reduction of the MORICS data. 
D.X., F.A.B. and R.C. contributed to the theoretical interpretation. 
L.D., J.W., and Y.A. contributed the analysis of the X-ray observations. 
R.C., Z.Z., A.Z., E.G.G.-M., M.L., Y.L., X.M., S.W., R.W., Y.W. and F.Y. contributed to the scientific interpretation and writing. 
All authors discussed the results and commented on the manuscript.

{\bf Competing interests:}
The authors declare that they have no competing interests.

{\bf Data and materials availability:}
The Keck/KCWI, HST, Subaru/MOIRCS, ALMA, VLA, and Chandra/ACIS data used in this work are publicly available. 
The data reported in this paper are available through the Keck/KCWI archive
(https://koa.ipac.caltech.edu/cgi-bin/KOA/nph-KOAlogin) with Program ID of N052, the ALMA and VLA archive (https://data.nrao.edu/portal/\#/) with Project Code of 17A-174 and 2018.1.00859.S, the HST archive (https://archive.stsci.edu/hst/) with the Program ID of 14760, the Chandra archive (https://cda.harvard.edu/chaser/) with the Program ID of 20357. 
The Subaru data can be found with the link of https://smoka.nao.ac.jp/fssearch? 
resolver=NONE\&object=BOSS1441\&instruments=MCS\&spectrographs=MCS\&obs\url{_}mod=all\&data
\url{_}typ=OBJECT\&dispcol=default\&action=Search\&obs\url{_}cat=all\&diff=1000\&asciitable=Table.



{\bf Supplementary Material:} 
Materials and Methods, Supplementary Text, Figures S1-S13, Tables S1-S4, and References {\it (32-78)}


\clearpage

\begin{figure}[!t]
\centering
\includegraphics[width=\textwidth]{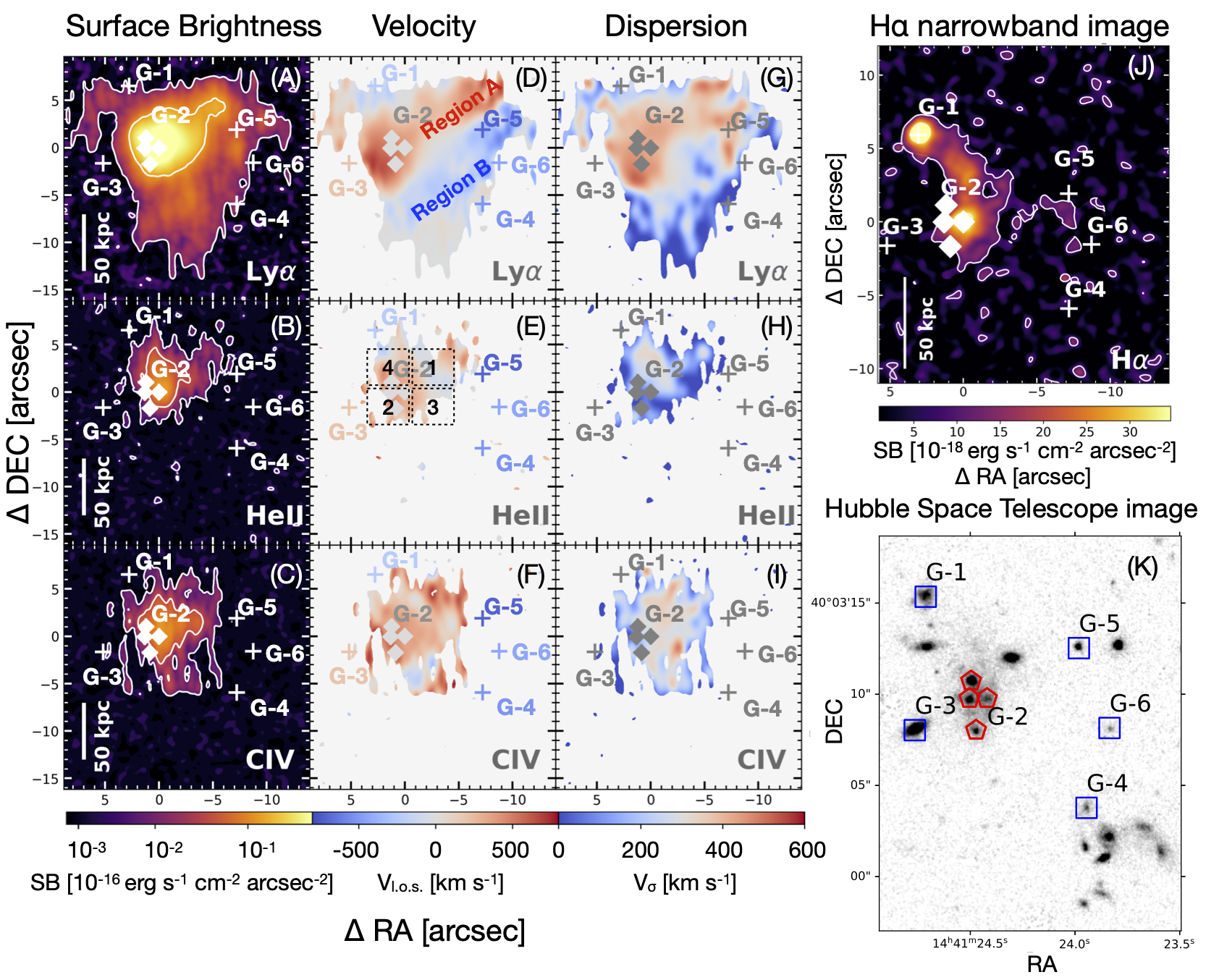}
\end{figure}
\noindent {\bf Figure 1: Images of the extended line emission.} {\bf (A to C):} Optimally extracted surface brightness (SB) images of Ly$\alpha$, \heii, and \civ\ emission from the KCWI datacube. 
$\Delta$ Right ascension (RA) and $\Delta$ Declination (DEC) are the coordinate relative to the position of G-2 (J2000 $14^{\rm h}41^{\rm m}24.42^{\rm s}$, $+40^{\circ}03'09.7''$).
Cross and diamonds labelled G-1 to G-6 galaxies at $z=2.31$, measured in both the CO (J=1→0) \cite{Emonts2019} and CO (J=3→2) observations \cite{Li2021}. 
White contours are at 2-$\sigma$ and 10-$\sigma$ detection of Ly$\alpha$ emission, and regions of 2-$\sigma$ and 8-$\sigma$  detection for \heii \ and \civ. 
In each case, the emission line extends over CGM scales.
{\bf (D to F):} Flux-weighted velocity maps. 
$v_{\rm l.o.s.}$ is the line-of-sight velocity.
The Ly$\alpha$ velocity map is labelled to indicate the large-scale red strip (region A) and blue strip (region B) (also see Fig.~4A). 
The four dashed squares in the \heii \ velocity map are each 4$''$ by 4$''$ wide, and the extracted 1-D spectra from each of the four apertures are shown in Fig.~2. 
{\bf (G to I):} Flux-weighted velocity dispersion maps. 
$v_{\sigma}$ denotes the velocity dispersion.
{\bf (J):} The H$\alpha$ narrowband image (seeing $0.4''$). 
Symbols are the same as in (A) to (C). 
The flux peak of H$\alpha$ coincides with the Ly$\alpha$, \heii, and \civ \ emission. 
{\bf (K):} Hubble Space Telescope image of the galaxy group, using the rest-frame optical filter, F160W.
Four clumps within $2''$ of G-2 are circled in red, and galaxies at  $z\approx2.3$ are marked with blue boxes. 
Coordinates are given in J2000 equinox.

\newpage
\begin{figure}[!t]
\centering
\includegraphics[width=\textwidth]{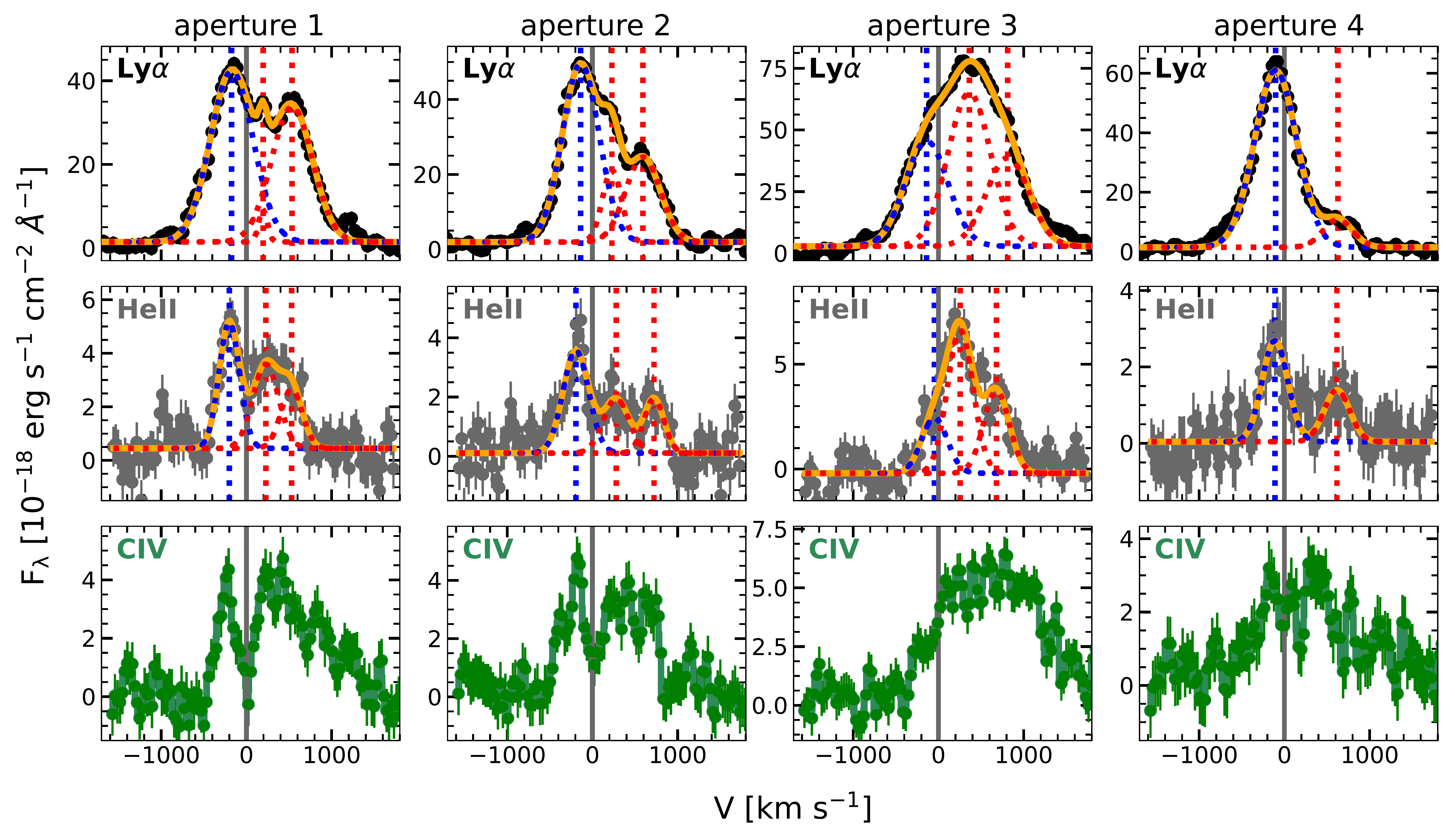}
\end{figure}
\noindent {\bf Figure 2: One-dimensional spectra of the Ly$\alpha$, \heii, and \civ \ emission.} 
These spectra were extracted from the apertures shown in Fig.~1E. 
The $y$-axis, $\rm F_{\lambda}$, denotes the flux density at wavelength of $\lambda$.
Dots with error bars show the observed spectra, while the orange lines are a model consisting of multiple Gaussians fitted to the data. 
The error bar of the dot represents the 1-$\sigma$ scatter which indicates 68\% confidence range. 
The individual Gaussian components are shown with dashed lines. 
 Numerical values are listed in table S1.
The vertical dashed lines mark the peaks of these Gaussian components. 
Each model requires two or three components to fit the data. 
Because \heii \ is a nonresonant line, the multiple components indicate gas motion of the CGM rather than resonant scattering effects \cite{Yang2014}. 
\civ\ is a doublet, so it was not modeled with Gaussians.


\newpage
\begin{figure}[b!t]
\centering
\includegraphics[width=\textwidth]{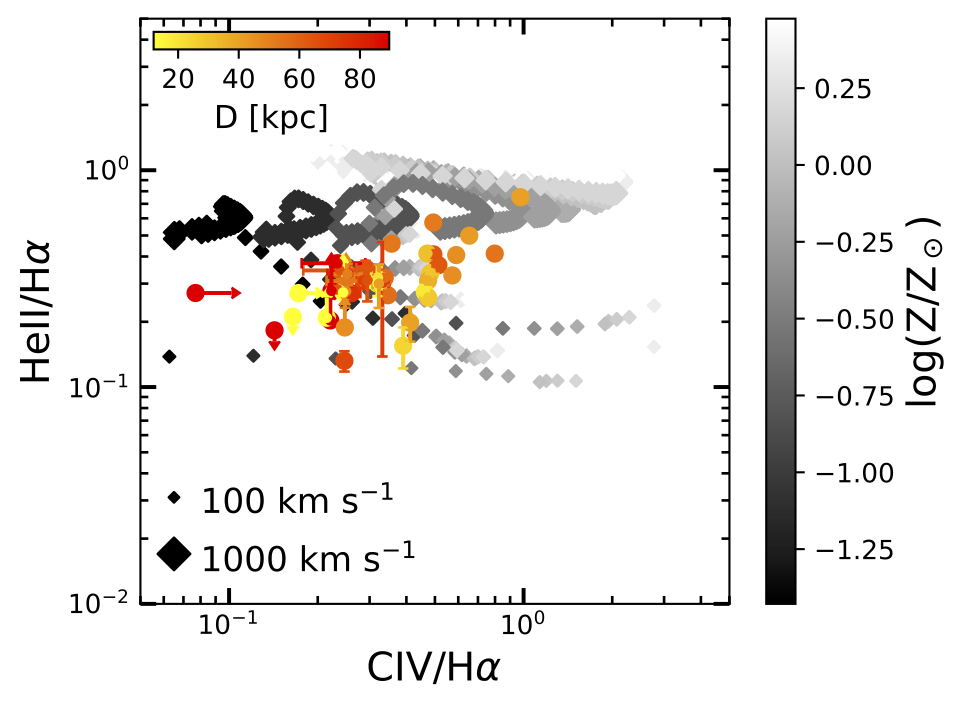}
\end{figure}
\noindent {\bf Figure 3: Line ratios \heii/H$\alpha$ and \civ/H$\alpha$ of 
the observations and the shock-with-precursor scenario \cite{Allen2008}. }
Color dots are the observations, and gray diamonds are predictions from the model \cite{Allen2008}. 
The error bar represents the 2-$\sigma$ scatter which indicates 95\% confidence range. 
The up (down) arrow denotes the lower (upper) limit.
The color encodes the projected distance (D) between the southeast aperture and each of the other apertures used to extract line emission (fig.~S4). 
The grayscale of the model points represents metallicity ($Z$), while their size scales with the shock velocity (100 --1000 km s$^{-1}$ in steps of 25 km s$^{-1}$), as indicated in the legend.
The observed line ratios of \heii/H$\alpha$ and \civ/H$\alpha$ are consistent with the shock-with-precursor scenario at the CGM metallicity of 0.1 to 1.0 $Z_{\odot}$ on a scale of $\approx$ 100 kpc.

\newpage
\begin{figure}[b!t]
\centering
\includegraphics[width=\textwidth]{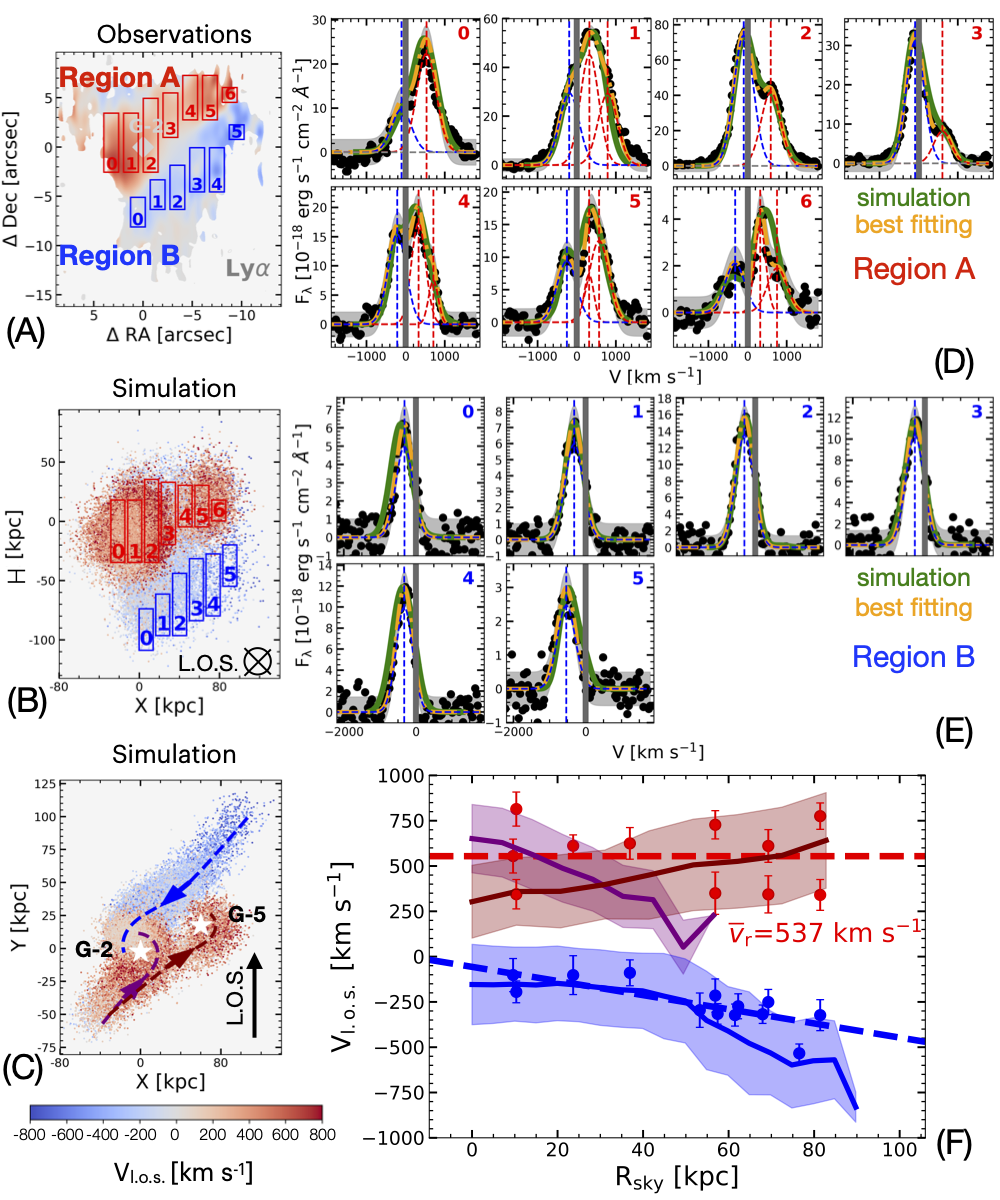}
\end{figure}
\noindent {\bf Figure 4: Interpretation as an inspiraling stream.} 
{\bf (A):} Same as Fig.~1D.
The red strip (region A) and blue strip (region B) are divided into subregions indicated by the boxes, within which the spectra in  (D) and (E) were extracted.
{\bf (B):} Same as  (A), but for our simulation of the inspiraling streams model.
The line of sight (L.O.S.) is along the $y$-axis. 
{\bf (C):} Projection of our model on the $x-y$ plane.
The two white stars mark the positions of G-2 and G-5. 
Three main inspiraling streams are shown by dashed lines, with flow directions indicated by arrows.
Two streams are around G-2 and the third stream is around G-5. 
Other projections of this model are shown in fig.~S9. 
{\bf (D and E):} Observed 1D spectra extracted from the regions shown in (A). 
The orange lines include multiple Gaussians, with each individual component shown as dashed lines as in Fig.~2. 
Black dots represent the data and the gray shading indicates the 1-$\sigma$ scatter.
The green lines are the simulated spectra extracted from the regions shown in (B). 
The data and simulation are consistent.
{\bf (F):} Projected line-of-sight velocity profile of the observations (dots) obtained from the red and blue components of the spectra in (D) and (E).
$\rm R_{sky}$ is the distance from each region to G-2 on the plane of the sky. 
The 11 red points correspond to the 11 redshifted Gaussians shown in (D), and the 13 blue points correspond to the 13 blueshifted Gaussians shown in (D) and (E). 
Error bars are 1-$\sigma$. 
The dashed lines are linear models fitted to the data points. 
The solid lines with shaded areas are our simulated velocity profiles, with colors corresponding to the streams in (C). 
The shaded areas denote the 1-$\sigma$ scatter.
The red components have a roughly constant velocity profile, whereas the absolute values of the blue components have an increasing profile. 
$\overline{v}_{\rm r}$ is the mean of the observed redshifted velocities.

\newpage
\setcounter{page}{1}

\begin{figure}
\centering
\includegraphics[width=.3\textwidth]{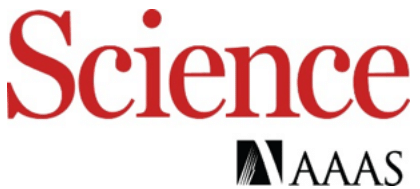}
\end{figure}

\begin{center}
    {\Large Supplementary Materials for}
    \section*{Inspiraling streams of enriched gas observed around a massive galaxy 11 billion years ago} 

    \author
{
{\renewcommand*{\thefootnote}{\fnsymbol{footnote}}
\setcounter{footnote}{1}
Shiwu Zhang,\footnotemark \ \setcounter{footnote}{1} Zheng Cai,$^{\ast}$\footnote{These authors contributed equally to this work} \ Dandan Xu, Rhythm Shimakawa,\\ 
Fabrizio Arrigoni Battaia,
Jason Xavier Prochaska,  Renyue Cen, \\ 
Zheng Zheng, Yunjing Wu, Qiong Li, Liming Dou, Jianfeng Wu,\\ Ann Zabludoff, Xiaohui Fan, Yanli Ai,  Emmet Gabriel Golden-Marx, Miao Li,\\
Youjun Lu, Xiangcheng Ma, Sen Wang, Ran Wang, Feng Yuan\\

}
}
\end{center}

~\\
{\bf The PDF file includes:}
\par Materials and Methods 
\par Supplementary Text 
\par Figs. S1 - S13 
\par Tabs. S1 - S4 
\par References (32-78) 
\newpage

\subsection*{Materials and Methods}
\def\theequation{S\arabic{equation}}

{\bf Observations.}
We obtained data with the KCWI \cite{Morrissey2018}, a blue-sensitive integral field unit instrument mounted on the  Nasmyth focus of the Keck II 10-m telescope. 
The large image slicer with a field-of-view (FoV) of 33$''\times$20$''$ was employed for the observations. 
Along the slicer direction, the spatial resolution is 1.3$''$.  Perpendicular to the slicer, the spatial resolution is 0.8$''$.
The KCWI/BM grating was used with two central wavelength settings, 
our observations cover the wavelength range of 3500 -- 5500 \AA. 
These configurations cover the redshifted Ly$\alpha$, \civ, \ and \heii \ emission. 
The resolving power was $R\approx4000$, limited by the slit width, corresponding to a velocity resolution of 75 km s$^{-1}$ (rest-frame 22.7 km s$^{-1}$). 

We performed a 2-hour exposure on the source and a 2-hour exposure on the sky in the Ly$\alpha$ region. 
This yields a 2-$\sigma$ surface brightness (SB) limit of 6.4$\times$10$^{-19}$ erg s$^{-1}$ cm$^{-2}$ arcsec$^{-2}$ at $\lambda\approx4100$ \AA, assuming a wavelength bin of 1 \AA.
For the \civ \ and \heii \ setting, the exposure time is also 2-hours on the source and 2-hours on the sky, giving a 2-$\sigma$ SB limit of 4.7$\times$10$^{-19}$ erg s$^{-1}$ cm$^{-2}$ arcsec$^{-2}$ at 5300 \AA \ in a wavelength bin of 1 \AA.

We used Multi-Object InfraRed Camera and Spectrograph (MOIRCS) \cite{Suzuki2008} on the Subaru telescope to observe the H$\alpha$ emission with narrowband and broadband imaging. 
MOIRCS covers 0.9 $\mu$m$-$2.5 $\mu$m and has a FoV of 4$'\times$7$'$. 
We used the narrowband filter BrG, which has a central wavelength $\lambda_{\rm c}=2.165$ $\mu$m and a bandwidth of $\Delta \lambda=0.025 \ \mu$m, appropriate for redshifted H$\alpha$ at $z=2.3$. 
The total exposure is 2.2-hours, yielding a 2-$\sigma$ SB limit of $9.4\times$10$^{-18}$ erg s$^{-1}$ cm$^{-2}$ arcsec$^{-2}$. 
We did our broadband observations using the K-band, which covers the same wavelength range as the of BrG narrowband. 
The integration time of the broadband was 30 minutes. {\color{black} The observing conditions for both the narrowband and broadband imaging were photometric with a median seeing of $\approx 0.4''$.} 

Previous radio observations \cite{Emonts2019,Li2021} acquired data for this field, covering the CO (J=1→0) and CO (J=3→2) emission lines (Fig.~S1A).
From those data, we measure the systemic redshift of MAMMOTH-1 to be $2.3116\pm 0.0004$ and $2.3120\pm 0.0006$ from the two CO lines, respectively.

We also proposed the X-ray observations for MAMMOTH-1 for this work (Fig.~S1B). 
We use the X-ray imaging instrument Advanced CCD Imaging Spectrometer (ACIS-I) \cite{Garmire2003} on the Chandra space observatory, which has a $0.5''$ per pixel spatial resolution.
The wide field imaging configuration was used with a field of view of 16$'\times$16$'$. 
{The observations have a total exposure time of 45 ksec.}
From the Chandra/ACIS observations, we identified a hard X-ray source at the location of G-2. 
The source count rate is 8 counts in the 2-8 keV band in an aperture with a radius of $3''$. 
We also measured that the background count rate corresponds to 1 count in the same aperture. 
The net count rate is estimated to be $1.4^{+0.8}_{-0.7} \times 10^{-4}$ cts/s by the Bayesian method. 
This X-ray source is detected by the \textsc{CIAO} \cite{CIAO2006} which is a software for analyzing the data from the Chandra space observatory. 
The false-positive probability threshold is $10^{-5}$ for the detection.
If we assume a simple power law including foreground Galactic hydrogen column density of $1.1\times10^{20}$ cm$^{-2}$ \cite{HI4PI-Collaboration2016}, the flux is estimated to be $5.1^{+6.9}_{-3.0}\times10^{-15}$ erg s$^{-1}$ cm$^{-2}$.
By fixing the photon index, 
we obtain a best-fitting column density of $3.5^{+3.2}_{-2.1}\times 10^{24}$ cm$^{-2}$, consistent with a Compton-thick active galactic nucleus (AGN) \cite{Gilli2007}.

{\bf Data reduction and pseudo-narrowband imaging.}
We adopt the standard KCWI pipeline \cite{KCWI_pipeline} to reduce our integral field spectrograph (IFS) data. For each image, we subtract the bias, correct the pixel-to-pixel variation 
with the flat-field images, and remove cosmic rays. 
We also conduct the geometric transformation and perform the wavelength calibration 
with ThAr arc images following a previously-published method \cite{Cai2019}. 
The datacube is constructed at this stage. Then, we use the twilight flats to correct the slice-to-slice variance and employ the spectroscopic standard star of Feige 34 to calibrate the flux of each individual image of the cube.

The pseudo narrowband images are obtained for the Ly$\alpha$, \heii, and \civ\ emissions with an optimal-extraction method \cite{Arrigoni2018b, Cai2019, Borisova2016}. 
Three sub-cubes are initially produced within the wavelength ranges where 
extended Ly$\alpha$, \heii, and \civ\ line emission is expected. 
The wavelength coverages of the three sub-cubes are fixed to be 4006${\rm \AA}$--4046${\rm \AA}$, 5411${\rm \AA}$--5451${\rm \AA}$, and 5109${\rm \AA}$--5149${\rm \AA}$, which cover each emission line.
To smooth the sub-cubes, we apply a Gaussian kernel with a full-width-at-half-maximum (FWHM) of $1''$, similar to the seeing.
We then construct the 3D segmentation masks containing values of 0 or 1 to select the connected voxels that have signal-to-noise, ${\rm S/N}\ge$ 2.
Such masks are used to extract weak extended line emission, and to obtain the higher-order moment maps. 

{\bf Systemic Redshift.}
To determine the flux-weighted velocity map, the systemic redshift of G-2 is required. 
We choose the redshift determined by the diffuse  CO (J=1→0) emission, $z=2.3116\pm0.0004$, as the systemic redshift of G-2 for three reasons. 
First, the radio observations show that the FWHM of the CO (J=1→0) emission of G-2 is 85 km s$^{-1}$ \cite{Emonts2019}.
This indicates that the molecular gas in the quasar host is dynamically colder than the ionized gas, so more likely to reflect the systemic redshift. 
The redshift obtained from CO (J=3→2), $z=2.3120\pm 0.0006$, is within 1-$\sigma$ of the result of the CO (J=1→0) measurement.
Second, previous work has shown that 
the dynamically cool CGM at large radii can be used to determine the systemic redshift of the host halo \cite{Cantalupo2012, Arrigoni2018a}. 
Our observations demonstrate that based on the redshift $z=2.3116\pm0.0004$, in the outskirts of the nebula which is $\ge 50$ kpc from G-2, the integrated Ly$\alpha$ emission has a central velocity of $5.3$ km s$^{-1}$.
This centroid has a 1-$\sigma$ uncertainty of 7.5 km s$^{-1}$. 
Thus, the diffuse Ly$\alpha$ in the outer regions of this nebula 
aligns with the systemic redshift determined by the diffuse CO (J=1→0) observations. 
Third, the flux-weighted redshift measured with the non-resonant line, \heii, is $z=2.3122\pm 0.0008$, also consistent with the result from CO (J=1→0). 
Therefore, we adopt a systemic redshift of  $z=2.3116\pm0.0004$.

{\bf Spectral Analysis.}
To extract and compare the Ly$\alpha$ and the \heii \ lines, we place four $4''\times4''$ apertures around the flux peak of the extended \heii.
We then fit a model consisting of multiple Gaussians to the spectra.
The best-fitting velocities are listed in Tab.~S1.
The Ly$\alpha$ velocities of the multiple Guassians are consistent with those of \heii \ within the 1-$\sigma$ range.
For a more detailed check, we use nine smaller apertures with $2.3''\times2.3''$ to map the \heii \ emission region with $\rm S/N\geq2$ (Fig.~S2). 
The resulting spectra are shown in Fig.~S2, and the best-fitting parameters are listed in Tab.~S2.
The Ly$\alpha$ emission has similar kinematics to \heii \ throughout the \heii \ emission region, indicating that the multiple components of Ly$\alpha$ arise primarily from the gas kinematics not radiative transfer (RT) effects.

Nevertheless, in Fig.~1, the flux-weighted velocities of \heii \ and Ly$\alpha$ 
appear to be different in some sub-regions. 
In these sub-regions, the flux ratio between the blue and red components of Ly$\alpha$ is different 
from that of \heii, but the multiple velocity components of Ly$\alpha$ and \heii\ still 
align with each other.  
To confirm this, we put a pseudo-slit on a sub-region where Ly$\alpha$ and \heii\ 
show different flux-weighted velocities.
In the 1-D spectra, the alignment of multiple velocity components between 
Ly$\alpha$ and \heii\ indicate that Ly$\alpha$ and \heii\ still trace the same cool gas (see Fig.~S2).

{\bf RT Effects Throughout the Nebula.}
We evaluate RT effects by plotting the SB profile and the radial profile of the velocity dispersion (dispersion profile) throughout the nebula. 
The SB profiles are shown on Fig.~S3A.
The SB profiles are fitted with an exponential function: 
\begin{equation}
    {\rm SB}(r)=Ce^{-R_{\rm sky}/r_{\rm h}},
    \label{SB_profile_fit}
\end{equation}
where $C$ is the normalization, $R_{\rm sky}$ is the radius to the source, and $r_{\rm h}$ is the scale length that describes the extent of the SB profile. 
We find scale lengths for the Ly$\alpha$ and \heii \ emission of $r_{\rm h,Ly\alpha}=17.8\pm0.8$ kpc and $r_{\rm h, HeII}=15.3\pm0.6$ kpc, respectively.
For the same flux level, the Ly$\alpha$ SB profile is as extended as that of the \heii \ emission.

The velocity dispersion was also used to evaluate RT effects. 
The line width of the resonant Ly$\alpha$ line could be larger than that of the \heii, if Ly$\alpha$ is dominated by RT effects \cite{Cantalupo2005, Yang2014}. 
The dispersion profiles are shown on the Fig.~S3B.
We fitted a linear model to the dispersion as a function of radius, finding that the \heii\ dispersion is almost constant, while the Ly$\alpha$ decreases at a higher radius.
The best-fitting slopes of the \heii \ and Ly$\alpha$ are $-0.1\pm0.4$ km s$^{-1}$ kpc$^{-1}$ and $-0.8\pm0.3$ km s$^{-1}$ kpc$^{-1}$, respectively. 
The dispersion ratio between Ly$\alpha$ and \heii \ decreases from $1.4$ to $1.1$ from small to large radii.
At larger radii, the impact of RT effects is weaker than at smaller radii, because the Ly$\alpha$ scattering strength primarily depends on the neutral hydrogen density \cite{Kimock2021}. 

At the small radii where RT effects cannot be neglected, 
the multiple components of Ly$\alpha$ and \heii \ align well with each other (Fig.~2, Tab.~S1 \& Tab.~S2).
These results indicate that, even in the inner region, the Ly$\alpha$ still traces the cool gas kinematics.
At larger radii, our analysis of SB and velocity dispersion profiles indicates that the RT effects are negligible, so Ly$\alpha$ should trace the cool gas kinematics.

The velocity dispersion shown in Fig.~S3B is measured with the individual Gaussians, obtained by fitting the spectra.
The spectra are extracted from the rectangular apertures of $1.4''\times1.4''$.
Since the dispersion shown in Fig.~S3B is measured with the individual Gaussians, it is consequently smaller than the line width shown in Fig.~S2 which contains two or three Gaussians.
The velocity dispersions of the individual Gaussians of Fig.~S2 are similar to the velocity dispersions shown in Fig.~S3B.

{\bf Line ratio diagnostics.}
Because H$\alpha$ is less sensitive to RT effects and less affected by dust than Ly$\alpha$, we calculate line ratios based on H$\alpha$ instead of Ly$\alpha$. We produce pseudo narrowband images of \heii\ and \civ\ from the data cubes with a bandwidth of 
$\Delta v=\pm$1700 km s$^{-1}$, equal to the width of the narrowband image of the H$\alpha$ emission. 
Then we calculate the line ratios of \civ\ and \heii\ emission relative to the H$\alpha$ emission. 
We use $1.5''\times1.5''$ apertures to extract the line ratios. In total, 44 apertures are used to calculate the line ratios and metallicities (Fig.~S4), with the measurements shown in Fig.~3. 
The southeast area of \heii \ emission is selected as the origin to calculate the projected distances of the apertures. 
The longest distance represents the projected physical scale of the diffuse metal-line emission.



We use the \textsc{CLOUDY (v17.02)} \cite{Ferland2017} to model the line ratios of the extended nebular emission. 
We follow previous work \cite{Cai2017a, Arrigoni2015a} in setting the parameters of \textsc{CLOUDY}. 
We select the built-in AGN continuum \cite{Mathews1987}.
We assume that the hydrogen volume density ($n_{\rm H}$) is constant and adopt standard plane-parallel geometry. 
We run \textsc{CLOUDY} simulations with parameters in the following ranges: $n_{\rm H}=$0.001 to 1 cm$^{-3}$ in a step of 1 dex, log $\frac{N_{\rm H}}{\rm cm^-2}=$17 to 21 in a step of 1 dex, where $N_{\rm H}$ is the column density of hydrogen, the ionization parameter log $U=$-3 to 0 in a step of 
0.1 dex, and the metallicity $Z=$0.01 to 5 $\times Z_{\odot}$ in a step of 0.6 dex, where $Z_{\odot}$ is the solar metallicity. 
The line ratios produced by pure photoionization modeling are shown in Fig.~S5. 
The average of the simulated \heii/H$\alpha$ is $2.6\pm0.5$, while the observed \heii/H$\alpha$ is $0.3\pm0.1$.
Therefore, the simulated line ratio of \heii/H$\alpha$ is one order of magnitude larger than the observations.

With the built-in AGN continuum with no dust included, the average predicted \heii/H$\alpha$ ratio is $2.6\pm0.5$. 
Nevertheless, both the AGN continuum and dust might influence the predicted line ratios.
Here, we consider a variety of possible AGN continua and include dust in the estimate of emission line ratios.
For the dust content, 
we use the built-in dust module in \textsc{CLOUDY}.
We set the dust type to interstellar medium (ISM), to approximate the CGM.
Under this setting, the ratio of extinction per reddening is $R_{\rm V}=3.1$, while the relation between the dust extinction in the V band and the hydrogen column density is $5.3\times10^{-22}(N_{\rm H}/{\rm cm}^{-2}) \ {\rm mag}$. 
Then, we construct a set of AGN continua by using an AGN continuum model \cite{Thomas2016}.
In this model, AGN continua are described by the energy of the peak of the UV bump ($E_{\rm peak}$), the photon index of the non-thermal emission ($\Gamma$), and the proportion of the total flux emitted in the non-thermal component ($p_{\rm NT}$).
Previous works \cite{Shang2005,Molina2013,Thomas2016} suggest the ranges of the parameters to be: $-2\leq{\rm log}(E_{\rm peak}/{\rm keV})\leq-1.8$ with a step of 0.08 dex, 
$1.7\leq\Gamma\leq3.5$ with a step of 0.45, 
and $0.1\leq p_{\rm NT}\leq 0.4$ with a step of 0.3.  
The parameters of \textsc{CLOUDY} are set to $n_{\rm H}=0.001$ to 1 cm$^{-3}$ in steps of 1.5 dex, log $\frac{N_{\rm H}}{\rm cm^-2}=$17 to 21 in steps of 2 dex, log $U=-3$ to 0 in steps of 
0.6 dex, and $Z/Z_{\odot}=$0.01 to 5 in steps of 1.4 dex.
These settings yield 5400 different models. 
All of AGN continua and 2D line ratio histograms are shown in Fig.~S6.
When the dust effects and the variety of the AGN continua are taken into account, the average \heii/H$\alpha$ is $2.3\pm0.8$. 
In the 5400 simulated models, only 17 models ($\sim0.3$\%) fall within 2-$\sigma$ of the observed values (Fig.~S6B).
These 17 models all correspond to $E_{\rm peak}=-2$ and $p_{\rm NT}=0.1$. 
Previous observations show that only 10\% of their AGNs have $E_{\rm peak}\leq 2$ \cite{Shang2005}, and only 8\% of their AGNs have $p_{\rm NT}\leq 0.1$ \cite{Jin2012}. 
Therefore, by taking dust and a variety of AGN continua into account, it is difficult to match the emission line ratios under the photoionization scenario.

In the shock-with-precursor scenario, we apply a library of radiative shock models, including the components of both the radiative shock and its photoionized precursor \cite{Allen2008}. In this model, the line emission is due to hard UV photons and soft-X-rays generated at the shock front induced by the fast winds. 
If the gas moves through the hot halo at a velocity higher than the local sound speed, a shock front forms at the leading edge of the cool gas clouds. 
Both the collisional ionization due to shocks and the photoionization produced by the soft X-ray precursor can power the line emission.
The flux of the line emission is mainly determined by the density of the precursor, $n_{\rm pre}$, the metallicity, $Z$, and the velocity of the fast shock, $v_{s}$. 
To cover the range of possibilities, 
the following parameter ranges are used: 1 cm$^{-3}\leq n_{\rm pre} \leq$1000 cm$^{-3}$ (steps of 1 dex), 
100 km s$^{-1} \leq v_{s} \leq$1000 km s$^{-1}$ (steps of 25 km s$^{-1}$), and 0.04$Z_{\odot} \leq Z \leq$2$Z_{\odot}$ (steps of 0.3 dex). 
The results, shown in Fig.~3, show that the shock with precursor model can reproduce the observed range of \heii/H$\alpha$ and \civ/H$\alpha$ line ratios. 
We do not use the spatial information to differentiate the powering mechanisms because the spatial resolution (11 kpc) is insufficient to differentiate shocks from photoionization effects. 

{\bf Dust Effects on Line Ratio Diagnostics.}
Our \textsc{CLOUDY} models including dust (see above) showed that the dust has a weak influence on the line ratios.
Here we estimate the line ratios under the assumption that the dust homogeneously mixes with the cool gas. 
Under this assumption \cite{Natta1984}:
\begin{equation}
    F^{\rm obs}_{\lambda}=F^{\rm int}_{\lambda}\frac{1-e^{-\tau_{\lambda}}}{\tau_{\lambda}},
    \label{dust_attenuation1}
\end{equation}
where $\tau_\lambda$ is the optical depth at a wavelength of $\lambda$, $F^{\rm obs}_{\lambda}$ is the observed flux at a wavelength of $\lambda$, 
and $F^{\rm int}_{\lambda}$ is the intrinsic flux at a wavelength $\lambda$.
By the definition of dust extinction, ${\rm A_{\lambda}/mag=2.5log(F^{\rm int}_{\lambda}/F^{\rm obs}_{\lambda})}$ \cite{Draine2011}, 
there is a relation between $\tau_{\lambda}$ and $A_{\lambda}$,
\begin{equation}
\tau_{\lambda}=\frac{A_{\lambda}}{\rm mag}/1.086,
\label{optical_depth1}
\end{equation}
where, by definition, $\tau_{\lambda}={\rm ln(F^{\rm int}_{\lambda}/F^{\rm obs}_{\lambda})}$.
By combining Eqs.~\ref{dust_attenuation1} and ~\ref{optical_depth1}, we have
\begin{equation}
    F^{\rm int}_{\lambda}=F^{\rm obs}_{\lambda}\frac{A_{\lambda}/1.086}{1-e^{-A_{\lambda}/1.086}}.
    \label{flux_correction}
\end{equation}
By applying the observed fluxes of the \heii \ $\rm 1640 \ \AA$, \civ \ $\rm 1548/1550 \ \AA$, and H$\alpha$ $\rm 6563 \ \AA$ emission lines, we can then determine the intrinsic flux of these lines.
We adopt the attenuation curve \cite{Salim2018}: 
\begin{equation}
    A_{\lambda}=\frac{A_{\rm V}k_{\lambda}}{R_{\rm V}},
    \label{attenuation_curve1}
\end{equation}
where $k_{\lambda}$ is the reddening curve \cite{Salim2018}, $A_{\rm V}$ is the dust extinction in the V band, and $R_{\rm V}=A_{\rm V}/(A_{\rm B}-A_{\rm V}) = 3.1$.
Under the assumption that the dust mixes well with the gas, we have $A_{\rm V}$ \cite{Draine2011}:
\begin{equation}
    A_{\rm V}=5.3\times10^{-22}(N_{\rm H}/{\rm cm}^{-2}) \ {\rm mag},
    \label{Extinction_V1}
\end{equation}
where the $N_{\rm H}$ is the total hydrogen column density. 
At $z=2$, the CGM around quasars has a total hydrogen column density of $10^{\rm 19.0-20.5}$ cm$^{-2}$ \cite{Lau2016, Prochaska2009, Hennawi2015, McCourt2018} with a host halo mass ranging from $10^{12.5}\ -10^{13.3} \ M_{\odot}$, bracketing the halo mass of the MAMMOTH-1 system. 
This hydrogen column density is consistent with the results from our \textsc{CLOUDY} models.
By adopting these values and using Eq.~\ref{Extinction_V1}, we have:  
\begin{equation}
    5.3\times10^{-3} \ {\rm mag}\leq A_{\rm V}\leq1.6\times10^{-1} \ {\rm mag}.
    \label{dust_extinction1}
\end{equation}

By combining Eqs.~\ref{attenuation_curve1} and ~\ref{dust_extinction1}, the dust extinction at $\lambda=1640$ \AA, $1549$ \AA, and $6563$ \AA\ is $0.022\leq A_{\rm 1640}\leq0.42$, $0.024\leq A_{\rm 1549}\leq0.43$, and $0.0035\leq A_{6563}\leq0.11$, respectively.
By substituting these values into Eq.~\ref{flux_correction}, we have $F_{1640}^{\rm int}=(1.01-1.22)F_{1640}^{\rm obs}$, $F_{1549}^{\rm int}=(1.01-1.22)F_{1549}^{\rm obs}$, and $F_{6563}^{\rm int}=(1.00-1.06)F_{6563}^{\rm obs}$. 
Therefore, the ratios between the intrinsic line ratios and the observed line ratios are $R_{\rm HeII/H\alpha}^{\rm int}\approx (1.01-1.15)R_{\rm HeII/H\alpha}^{\rm obs}$ and $R_{\rm CIV/H\alpha}^{\rm int}\approx (1.01-1.15)R_{\rm CIV/H\alpha}^{\rm obs}$. These values suggest that the dust in the MAMMOTH-1 nebula has up to a 15\% effect on the emission line ratios of \heii/H$\alpha$ or \civ/H$\alpha$.  
This is too small to affect our qualitative conclusions.

{\bf The Estimation of Star Formation Rate (SFR), Stellar Mass, and Halo Mass.}
The SFR of G-2 is calculated from the far-infrared (FIR) luminosity (rest-frame $8-1000 \ \mu$m), which is $L_{\rm FIR}=2.0\pm0.4 \times10^{45}$ erg s$^{-1}$ \cite{Li2021}. 
From the spectral energy distribution (SED) of G-2 with \textsc{Cigale} \cite{Boquien2019}, the AGN FIR luminosity is $L_{\rm AGN, FIR}=1.7 \times 10^{44}$ erg s$^{-1}$. Thus, the AGN accounts for approximately 8.5\% of the total FIR luminosity. 
By using the AGN-removed FIR luminosity of $L_{\rm FIR-AGN}=1.8\pm0.4\times10^{45}$ erg s$^{-1}$ 
and the relation ${\rm SFR(M_{\odot}/yr)} = 4.5\times10^{-44} (L_{\rm FIR}/{\rm erg \ s^{-1}})$ with the intrinsic uncertainty of about 0.1 dex \cite{Kennicutt1998}, we estimate SFR to be $81\pm 18 \ M_{\odot}$ yr$^{-1}$.

The halo mass of G-2 is obtained from the stellar mass-halo mass relation (SMHM) taken from the IllustrisTNG simulations at $z=2$ \cite{Nelson2019}.
We derive the stellar mass of G-2 from the SED: ${\rm log}(M_{\star}/M_{\odot})=11.1^{+0.2}_{-0.3}$. Using the SMHM relation, the host halo mass of G-2 is ${\rm log}(M_{\rm h}/M_{\odot})=12.9^{+0.4}_{-0.4}$,
consistent with the halo mass of ${\rm log}(M_{\rm h}/M_{\odot})=12.8^{+0.5}_{-0.4}$ derived by adopting an alternative empirical model of the SMHM \cite{Lu2014}.

Assuming that the system is virialized, the upper limit of the host halo mass of G-2 is \cite{Evrard2008}:
\begin{equation}
    M_{\rm h}=(10^{15}M_{\odot}/h(z))(\frac{\sigma_{v}}{1083 \ {\rm km\ s^{-1}}})^{1/0.34},
    \label{halo_mass}
\end{equation}
where $M_{\rm h}$ is the halo mass, $\sigma_{v}$ is the velocity dispersion of the halo, and $h(z)$ is the normalized Hubble parameter at the redshift of $z$.
Because our CO (J=3→2) observations provide redshift of the sources (Tab.~S3).
By applying this value to Eq.~\ref{halo_mass}, we find an upper limit on the host halo mass of G-2 of ${\rm log(M_{\rm h, up}/M_{\odot})}=13.4^{+0.1}_{-0.1}$.
This upper limit is also consistent with the value from the SMHM. 
We therefore adopt the host halo mass of G-2 to be ${\rm log}(M_{\rm h}/M_{\odot})=12.9^{+0.4}_{-0.4}$.

{\bf Metal-Enriched Inspiral CGM Kinematics.} 
We use the Illustris TNG cosmological simulations \cite{Nelson2019} to analyze the CGM kinematics.
We focus on the TNG-100 simulation box because it has a sufficient number of massive halos to statistically assess the kinematics and physical properties of the CGM.
This simulation has a baryon mass resolution of 1.4$\times$10$^{6}\ M_{\odot}$ and a gravitational softening length of 0.185 kpc, sufficient to the study large-scale gas kinematics \cite{Nelson2019}. 
G-2 has a host halo mass of ${\rm log}(M_{\rm h}/M_{\odot})=12.9^{+0.4}_{-0.4}$, ${\rm SFR}=81\pm 18 \ M_{\odot}$ yr$^{-1}$, and stellar mass of ${\rm log}(M_{\star}/M_{\odot})=11.1^{+0.2}_{-0.3}$.
To match these measured quantities {\color{black} and to have a sample size large enough for a statistical study}, we select galaxy group systems at $z=2$ from the simulation based on the following three criteria:
\begin{itemize}
\item The host halo mass of the main galaxy is larger than 2$\times$10$^{12} \ M_{\odot}$.
\item The stellar mass of the main galaxy is larger than 2$\times$10$^{10} \ M_{\odot}$.
\item The star formation rate of the main galaxy is larger than 5 $M_{\odot}$ yr$^{-1}$.
\end{itemize}
91 simulated galaxies pass these cuts in the TNG-100 box at $z\approx2$. 
The properties of the systems are shown in the Fig.~S7A.
To study the cool gas kinematics in the CGM, we limit the gas temperature range to 10$^{4- 5}$K, similar to the CGM gas that is emitting Ly$\alpha$, \civ, \ and \heii.

To study the kinematics of the CGM, we employ the spin parameter, $\lambda_{\rm cool}$, which describes the rotation of the cool gas in systems:
\begin{equation}
\lambda_{\rm cool}=\frac{j_{\rm cool}}{\sqrt{2}R_{\rm vir}V_{\rm vir}},
\label{spin_parameter}
\end{equation}
where $j$ is the specific angular momentum, and $R_{\rm vir}$ and $V_{\rm vir}$ are the virial radius and velocity of the dark matter halo, respectively. 
Previous studies \cite{Stewart2011, Danovich2015, Teklu2015, Stewart2017} have shown
that the cool gas in the CGM of cosmological simulations has a higher spin parameter ($\lambda_{\rm cool} \approx 0.1$) than the other components (dark matter, halo gas, and stars). 
To study the CGM gas metallicity ($Z$), the mean metallicity of the cool gas particles within the range of $0.1R_{\rm vir}$ to $50 \ {\rm kpc}$ are employed.
By adopting the criteria of $\lambda_{\rm cool}\geq 0.1$ and $Z\geq 0.25Z_{\odot}$, we find that 42 out of the 91 simulated systems have metal-enriched inspiraling streams with high angular momentum. 
This is consistent with previous work \cite{Stewart2011,Stewart2017}, and indicates that simulated CGM can be enriched to a few tenths $Z_{\odot}$ out to a radius of 50 kpc.



Of the 42 simulated systems, the majority have metal-enriched gas inflow (Fig.~S7C), 
for simulated massive halos with $M>10^{12}\ M_\odot$ at $z\approx2.0$.
There is no strong dependence of the presence of a recycled inflow on mass or environment. 
This is because CGM enrichment is mainly an internal process due to feedback from a starbust or AGN activity from the central galaxy.
The nearby satellite galaxies only bring angular momentum and perturb the cool gas \cite{Wang2021}. 
The example shown in Fig.~S8A is selected at $z=2$ from the TNG-100 box with the halo mass of $M_{\rm h}=10^{13.1} \ M_{\odot}$ and the $\rm SFR$ of $56.7 \ M_{\odot}$ yr$^{-1}$. 
This simulated system resides in a group environment, similar to MAMMOTH-1.

In TNG-100 simulations, 13 systems have halo masses with $M_{\rm{halo}}>10^{13}\ M_\odot$ at $z=2.0$. We find that six out of the 13 (46\%) massive systems ($M_{\rm h}\geq 10^{13} \ M_{\odot}$, Fig.~S7B) at $z=2.0$ have $\lambda_{\rm cool}\geq0.1$ and $Z\geq 0.25Z_{\odot}$. 
To provide better statistics, we enlarge the sample by including massive systems at $z=1.9$ in the TNG-100 box.
57 massive systems with the halo mass larger than 10$^{13} M_{\odot}$ are selected. The halo masses range from $1.0\times 10^{13}M_{\odot}$ to $5.8\times 10^{13}M_{\odot}$.
Among these halos, 24 systems (42\%) have cool gas with spin parameters and metallicities beyond $\lambda\geq0.1$ and $Z\geq0.25 \ Z_{\odot}$ consistent with the results at $z=2.0$. 
This is different from predictions of the cold-mode accretion, which find 
that inflowing gas is pristine with the metallicity of $\leq10^{-3} Z_\odot$ \cite{Martin2015}.

{\bf Geometry of Inspiraling Streams.}
Comparing the inspiraling streams in simulations with the arms of spiral galaxies, we find that they are structurally similar. Both show a tail and a spiral head.
An example is shown in Fig.~S8A.
Due to this similarity, we use the formula describing an arm \cite{Ringermacher2009}  of a spiral galaxy to construct the geometry:
\begin{equation}
r( \theta)=\frac{A}{{\rm ln}(B\cdot {\rm tan}(\omega \theta))},
\label{spiral_func0}
\end{equation}
where $r(\theta)$ is the radius to the central source, $A$ is the scale parameter, $B$ and $\omega$ determine the curvature, 
and $\theta$ controls the length of the spiral arms, $l$, which is the projection of the stream on its asymptotic line (Fig.~S9A). 
For this geometry, the line-of-sight is the $y$-axis, and the $x-z$ plane is the plane of the sky.  
To characterize the orientation of the streams, we introduce $\alpha$,  the angle between the streams and 
the $x$-axis. 
By accounting for this orientation, we modify Eq.~\ref{spiral_func0} to be:
\begin{equation}
r( \theta)=\frac{A}{{\rm ln}(B\cdot {\rm tan}(\omega (\theta-\alpha)+\beta))},
\label{spiral_func}
\end{equation}
where $\beta$ = arctan $B^{-1}$. 
We model the two inspiraling streams shown in Fig.~S8A  by fitting them with Eq.~\ref{spiral_func}, which yields the reduced chi-square values of $\chi_{\rm 01}^{2}=0.9$ and $\chi_{\rm 02}^{2}=0.7$, respectively.


To construct our geometric model, we introduce the thickness of the streams, $h$, the width of the streams, $w$, and the inclination of the stream, $i$.   
We place the following constraints on our model:

(i) The values of $A$, $B$, and $\omega$ are constrained by fitting the inspiraling streams of TNG galaxies described above with $M_{\rm h}\geq10^{13}\ M_\odot$ at $z=2$. 
The median of each of these parameters is $A=177$ kpc, $B=0.01$, $\omega=-0.25$. 

(ii) We restrict $32^{\rm o}\leq \alpha \leq 52^{\rm o}$. 
For the lower limit of $\alpha$, simulations show that the maximum velocity of the cool gas in the massive halos (${\rm log}(M_{\rm h}/M_{\odot})\ge13$) is $v_{\rm max,sim}=1470$ km s$^{-1}$. 
To produce the observed maximum  line-of-sight velocity which is $v_{\rm max, obs}=801$ km s$^{-1}$, $\alpha$ must be at least $\alpha_{\rm low}={\rm arcsin(801 \ km s^{-1}/1470 \ km s^{-1})}\approx 32^{\rm o}$. 
For the upper limit of $\alpha$, the projected length of the stream gets smaller with increasing $\alpha$. 
For halos of ${\rm log}(M_{\rm h}/M_{\odot})=13$, we adopt a halo radius of 200 kpc. The largest length of the inspiraling streams must be no more than the radius of the halo.
The observations show that the projected physical length of the longest stream, that is, stream-1, is 124 kpc (blue aperture 0-5 in Fig.~4A).
To produce the projected length of the inspiraling stream, the angle must be no more than $\alpha_{\rm up}={\rm arccos(124 \ kpc/200 \ kpc})\approx 52^{\rm o}$.
We then take $\alpha=42^{\rm{o}}$, within the range of $32^{\rm o}\leq \alpha \leq 52^{\rm o}$, to construct MAMMOTH-1 model.

(iii) $\theta$ is restricted to be in the range $\theta_{\rm low}\leq \theta \leq \alpha+\pi$, where $\theta_{\rm low}$ is larger than $\alpha$. 
The length of the stream, $l$, is related to $\theta_{\rm low}$.
\begin{equation}
    l=|r(\theta_{\rm low}){\rm cos}(\theta_{\rm low}-\alpha)|+|r(\alpha+\pi)|.
    \label{spiral_length}
\end{equation}
Given $\alpha=42^{\rm o}$, the length of the stream can be obtained from its projection ($l_{\rm pro}$) on the sky with $l=l_{\rm pro}/{\rm cos}(\alpha)$. 
From Fig.~4A, we take the distance from blue aperture 0 to 5 as the projection of stream-1 ($l_{\rm pro,1}=124$ kpc), the distance from red aperture 0-2 as the projected length of stream-2 ($l_{\rm pro,2}=48$ kpc), and the distance from the red aperture 1 to 6 as the projected length of stream-3 ($l_{\rm pro,3}=106$ kpc).
These projected lengths yield $l_{1}=166$ kpc, $l_{2}=65$ kpc, and $l_{3}=142$ kpc.
By applying these lengths to Eq.~\ref{spiral_length}, we have $\theta_{\rm low, 1}=1.24\pi$, $\theta_{\rm low, 2}=0.27\pi$, and $\theta_{\rm low, 3}=0.25\pi$.

(iv) The thicknesses of the two red streams ($h_{2}=55$ kpc and $h_{3}=55$ kpc) are determined by the width of Region A, while the thickness of the blue stream ($h_{1}=120$ kpc) is determined from the width of the Ly$\alpha$ nebula. The geometry is shown in Fig.~S9C.

(v) We assume the inclination to be $i=90^{\rm o}$, that is, that the inspiraling streams are edge-on. This parameter does not have to be $i=90^{\rm o}$ exactly. 
Rather, the model allows the range $70^{\rm o}\leq i \leq 110^{\rm o}$ (see discussion below).

(vi) Because we do not know the distance along the line of sight, the width of the streams is assumed to be $W=55$ kpc.
This parameter only influences the line-of-sight positions of the simulated gas particles, not the line-of-sight velocities or the gas particle distributions on sky. 
Thus, the width of the streams does not influence the simulated spectra.

{\bf Kinematics of Inspiraling Streams.}
To model the velocity along the streams, $v_{\rm stream}$, a linear function is implemented. 
This is because the velocity profile from simulations can be explained with a linear profile (Fig.~S8B), 
and the velocity profile can be described as
\begin{equation}
v_{\rm stream}=k_{v}r+v_{b},
\label{vprofile}
\end{equation}
where $r$ is the distance to the central source, $k_{v}$ is the slope of the linear profile, and $v_{b}$ is the intercept of the linear profile. 
To simulate the velocity dispersion, we add a Gaussian distribution to the velocity 
\begin{equation}
v_{\rm stream}=k_{v}r+v_{b}+\delta v_{\rm stream},
\end{equation}
where $\delta v_{\rm stream}$ follows a normal distribution, $G_{n}(0,\sigma_{v})$, with $\sigma_{v}$ as the velocity dispersion. Then, the velocity along the line 
of sight can be described as:  
\begin{equation}
v_{\rm l.o.s.}=(k_{v}r+v_{b}+\delta v_{\rm stream})\cdot \sin(\phi),
\end{equation}
where the $v_{\rm l.o.s.}$ is the line-of-sight velocity of the gas particle, and $\phi$ is the angle between the gas velocities and the sky plane (Fig.~S9A\&B).

The geometric parameters are all fixed by comparing them to observations or simulations (Tab.~S4).
The kinematic parameters are free parameters, of which there are nine: $k_{v}$, $v_{b}$, and $\sigma_{v}$ for each of the three streams.
To determine the best-fitting values, we minimize the residual between the observed spectra and the simulated spectra from the same aperture (Fig.~4D \& E) and apply a Markov Chain Monte Carlo (MCMC) analysis. 
We initialize this process with thirty Markov chains and iterate each chain over 100,000 steps.
During burn-in to throw away the first steps to initialize the iteration, the first 1000 steps are discarded. 
The acceptance rate is set to 0.25. 
The output parameters are shown in Tab.~S4 and Fig.~S10. 

{\bf Robustness of Inspiraling-stream Model.}
The flux-weighted velocity map of MAMMOTH-1 shows a red strip and blue strip (Fig.~1D). 
In the red strip, the nebula shows double or triple peaks in velocity. 
In our model, the multiple peaks in the red strip are due to two metal-enriched inspiraling streams. 
Our model is consistent with the data shown in Fig.~4D\&E. 

Nevertheless, the specific geometric configuration shown in Fig.~4C is not the only solution. 
The synthetic spectra derived from our model are not sensitive to the geometric configurations within a certain range. 
For stream-1, we tested different orientations ($\alpha=32^{\rm o}, 52^{\rm o}$), shape parameters ($\omega=-0.5, -0.125$) and ($B=0.005,0.02$), and inclinations ($i=70^{\rm o}, 110^{\rm o}$).
The results (Fig.~S11) show that even though the configurations are different, the model yields the almost the same line-of-sight velocity profile ($\chi_{\rm r}^{2}\approx 0.8$). 
This shows that the qualitative form of our model is robust, but its quantitative parameters are not well constrained by the data.

{\bf Average Mass Inflow Rate Estimation.}
The construction of the metal-enriched inspiraling-stream model allows us to estimate the mass inflow rate \cite{Fox2019}. 
 The mass inflow rate is defined as:
\begin{equation}
\dot{M}_{\rm in}=\frac{\Delta M_{\rm in}}{\Delta r}\cdot v_{\rm r},
\label{inflowrate}
\end{equation}
where $\Delta M_{\rm in}$ is the mass of the inflowing gas, $\Delta r=50$ kpc is the radius to the central source (G-2), and $v_{\rm r}$ is the average radial velocity. 
$\Delta M_{\rm in}$ is calculated with Eq.~\ref{Mciv}, which gives $\Delta M_{\rm in}=(1.9\pm 0.4)\times 10^{11} \ M_{\odot}$.

$v_{\rm r}$ is calculated from the average radial velocities within annulus. 
The annulus has an outter radius of 50 kpc and an inner radius of 20 kpc centering on G-2.
From our models, we find $v_{\rm r}=181^{+26}_{-20}$ km s$^{-1}$.  
With the radial velocity and gas mass, we calculate $\dot{M}_{
\rm in}=703^{+101}_{-78} \ M_{\odot}$ yr$^{-1}$ where the uncertainty is the 16\% to 84\% percentile range.



\subsection*{Supplementary Text}

{\bf Outflow kinematics.}
The large-scale kinematics we observe are inconsistent with an ongoing outflow. 
First, we expect an ongoing outflow to decelerate at large projected distances.  
This deceleration yields a decreasing line-of-sight velocity profile, because the outflow continues to lose kinetic energy due to the gravitational potential of the dark matter halo and the pressure of the surrounding medium \cite{Fabian2012,Richings2018}.
Under the assumption that the outflow velocities are uniformly distributed within the angle between the sightline and the outflow direction, projection effects also contribute to the decreasing velocity.
Contrary to the results expected from an ongoing outflow, the observed line-of-sight velocity profile (Fig.~4F) indicates that the redshifted velocity is roughly constant and that the absolute value of the blueshifted velocity increases with increasing distance from G-2.



Second, modeling the data as an outflow requires a very high value of the coupling efficiency ($f_{c}$). 
The coupling efficiency is the ratio between outflow power and the AGN bolometric luminosity, so describes how efficiently the AGN bolometric luminosity couples with the gas in the ISM or CGM.
To calculate $f_{c}$, we employ two models.
The first model assumes that radiative cooling is inefficient, so that energy is conserved on large scales \cite{Cai2017a, Nesvadba2008, Harrison2012, Harrison2014}. 
 Under this assumption, the outflow energy rate is $\dot{E}_{\rm out}=1.5\times 10^{46} {R_{\rm   out}}^{2}{v_{\rm out}}^{3}n_{e}$ erg s$^{-1}$, 
where $R_{\rm out}$ is the outflow radius in units of kpc, $v_{\rm out}$ is the outflow velocity in units of km s$^{-1}$, and $n_{e}$ is the 
electron density in units of cm$^{-3}$. 
Because energy is conserved, it can be regarded as the upper bound of the outflow energy rate. 
By adopting our observational results, $\rm R_{out}=54$ (half of the spatial extent of \civ \ emission), $v_{\rm out}=537$ (the constant red components in Fig.~4F), and assuming electron density of $n_{e}=1.5$, we calculate that this upper bound is $\rm \dot{E}_{\rm out, up}=2.03\times10^{47}$ erg s$^{-1}$.

The second model gives the lower bound on the outflow energy rate \cite{Cano-Diaz2012, Greene2012, Harrison2014, Herrera-Camus2019}. It assumes that the outflowing gas 
extends from 0 kpc to the radius where we see the extended emission $R_{\rm out}$. 
Thus, the outflow mass rate ($\dot{M}_{\rm out}$) is given by the outflow gas mass time-averaged over the flow timescale $\dot{M}_{\rm out}=M_{\rm out}v_{\rm out}/R_{\rm out}$, where $M_{\rm out}$ is the outflow mass, $v_{\rm out}$ is the outflow velocity, and $R_{\rm out}$ is the radius of the outflow.
The luminosity of \civ \ emission can be converted to the outflow gas mass, $M_{\rm out}$ \cite{Cano-Diaz2012}:
\begin{equation}
M_{\rm out}=\frac{1.27m_{\rm H}n_{e}L_{\rm CIV}}{\epsilon_{\rm CIV}},
\label{Mciv}
\end{equation} 
where $m_{\rm H}$ is the mass of a Hydrogen atom, $n_{e}$ is the electron number density, and $L_{\rm CIV}$ and $\epsilon_{\rm CIV}$ are the luminosity and emissivity of \civ \ emission \cite{Pequignot1991,Cano-Diaz2012}, respectively. 
By applying $n_{e}=1.50$ cm$^{-3}$ and $L_{\rm CIV}=10.50\times 10^{42}$ erg s$^{-1}$ in Eq.~\ref{Mciv}, we obtain $M_{\rm out}=6.41\times$10$^{11}{\rm M_{\odot}}$. 
Using the formula, $\dot{E}_{\rm out}=\frac{1}{2}\dot{M}_{\rm out}(v_{\rm out}^{2}+3D^{2})$, we calculate the outflow energy rate \cite{Herrera-Camus2019, Cano-Diaz2012, Harrison2014}. 
By applying the mass outflow rate, we have $\dot{E}_{\rm out}=\frac{1}{2}\frac{M_{\rm out}{v_{\rm out}}}{R_{\rm out}}(v_{\rm out}^{2}+3D^{2})$. 
With the observed velocity ($v_{\rm out}=537$ km s$^{-1}$), radius ($R_{\rm out}=54$ kpc), and velocity dispersion ($D=258$ km s$^{-1}$),
we obtain $\dot{E}_{\rm out}=9.80\times10^{44}$ erg s$^{-1}$. 
For the final outflow energy, we adopt the mean of the upper and lower bound in log space \cite{Harrison2012, Harrison2014}, which gives $\dot{E}_{\rm out}=1.41\times10^{46}$ erg s$^{-1}$. 
With the available multi-wavelength observations \cite{Cai2017a,Emonts2019}, we model the SED and estimate the bolometric luminosity of G-2 using \textsc{Cigale} \cite{Boquien2019} (Fig.~S12). 
We find the AGN bolometric luminosity is $L_{\rm bol}=2.42\times 10^{45}$ erg s$^{-1}$.
By applying $L_{\rm bol}$  to calculate the coupling efficiency, we find $f_{\rm c}=5.82$. 
This value is almost one order of magnitude larger than previous observations and predictions from simulations (Fig.~S13) \cite{Harrison2018}. 

{\bf Intergalactic Transfer between Galaxies.}  Observations show that the Ly$\alpha$ emission peaks at G-2. The neighbouring galaxies (G-1, G-3 -- G-6) reside at the edge of this extended emission. Thus, the diffuse
emission mostly traces the circumgalactic gas of G-2. 
If the cool gas traced by the Ly$\alpha$ emission is the intragroup medium originating from intergalactic transfer from the satellite galaxies, 
then the metal-enriched gas would be likely to cover the neighbouring galaxies.
This is because the cool gas transferred from the ISM is enriched at $\geq 0.1  Z_{\odot}$. 
However, our observations suggest no metal emission detected around the neighboring galaxies.
Simulations predict that intergalactic transfer provides $\leq 10\%$ of the stellar mass until $z=2$ \cite{Angles-Alcazar2017}. 
Thus, both observations and simulations suggest that the accretion of the recycled CGM gas is the dominant process, which is also consistent with the kinematics (Fig 2).

Tidal stripping could also be a mechanism of metal enrichment and intergalactic gas transfer between galaxies. 
This effect becomes non-negligible once the distance between the central and the satellite galaxies is smaller than Roche radius:
\begin{equation}
    R_{\rm r}=r_{s}\cdot (\frac{2M_{c}}{m_{s}})^{1/3},
    \label{tidal_stripping}
\end{equation}
where $r_{s}$ is the radius of the satellite, $M_{c}$ is the dynamical mass of the central source enclosed within $R_{\rm r}$, and $m_{s}$ is the dynamical mass of the satellite galaxy.
G-3 is taken as an example to illustrate tidal stripping.
$r_{s}$ is taken as the half-light radius of G-3, which is $r_{s}=3$ kpc from the Hubble Space Telescope image.
$m_{s}$ is the total mass within 3 kpc, which is dominated by the stellar mass.
The stellar mass of G-3 is estimated to be $2.5\times10^{10} \ M_{\odot}$ converted from the H$_{2}$ gas mass with a $\rm H_{2}$-stellar-mass conversion \cite{Popping2014} shown in Tab.~S3. 
$M_{c}$ is set to the halo mass of G-2 ($10^{12.9} \ M_{\odot}$). 
Since the mass included within the radius of 100 kpc is much smaller than the halo mass, this setting gives an upper limit of the Roche radius.
By applying these values to Eq.~\ref{tidal_stripping}, we obtain $R_{\rm r, G-3}\le 27$ kpc, roughly two times smaller than its projected distance to G-2 (49 kpc). 
Because the other satellites are further away from G-2, the tidal forces are weaker. 
Therefore, we conclude the tidal stripping effect is estimated to be small in MAMMOTH-1 on CGM scales. 
Note that the estimation above is based on current observations. 
Satellite galaxies could have been closer to G-2 in the evolution history.

If we consider that all sources visible in the Hubble Space Telescope image are members of this galaxy group, the interaction between galaxies due to tidal force is still too weak to strip the ISM at large scale. 
Nevertheless, nearby satellite galaxies can influence the observed kinematics directly, by transferring angular momentum to the cool gas of the CGM.
If the cool gas is mainly around G-2 which has a halo mass of $M_{\rm h}\approx10^{13} \ M_{\odot}$, the enriched cool gas flowing onto G-2 is more likely to be the dominating process in this system.

{\bf Comparison with Previous Results.} 
Previous work \cite{Cai2017a} observed MAMMOTH-1 using narrowband imaging and slit spectroscopy.  
The length of the slit was only about $10''$, with only one position angle (PA). 
No sub-mm data or systemic redshift were available. 
That study interpreted the metal-enriched CGM as an outflow, but was based on more limited information.

The archival radio observations \cite{Emonts2019} have provided the systemic redshift of MAMMOTH-1. 
Our observations have provided the kinematics of the CGM. 
By comparing the spatially-resolved spectra of Ly$\alpha$ and \heii\ across 
the CGM, 
we find that the velocity structure of Ly$\alpha$ is not consistent with an 
AGN outflow model (see Fig.~4F and discussion above). 
If the extended emission is powered by AGN feedback, then it would require a high coupling efficiency, 
much larger than previous observations and simulations. 
We therefore reject this interpretation.



\newpage
\begin{figure}[b!t]
\centering
\includegraphics[width=\textwidth]{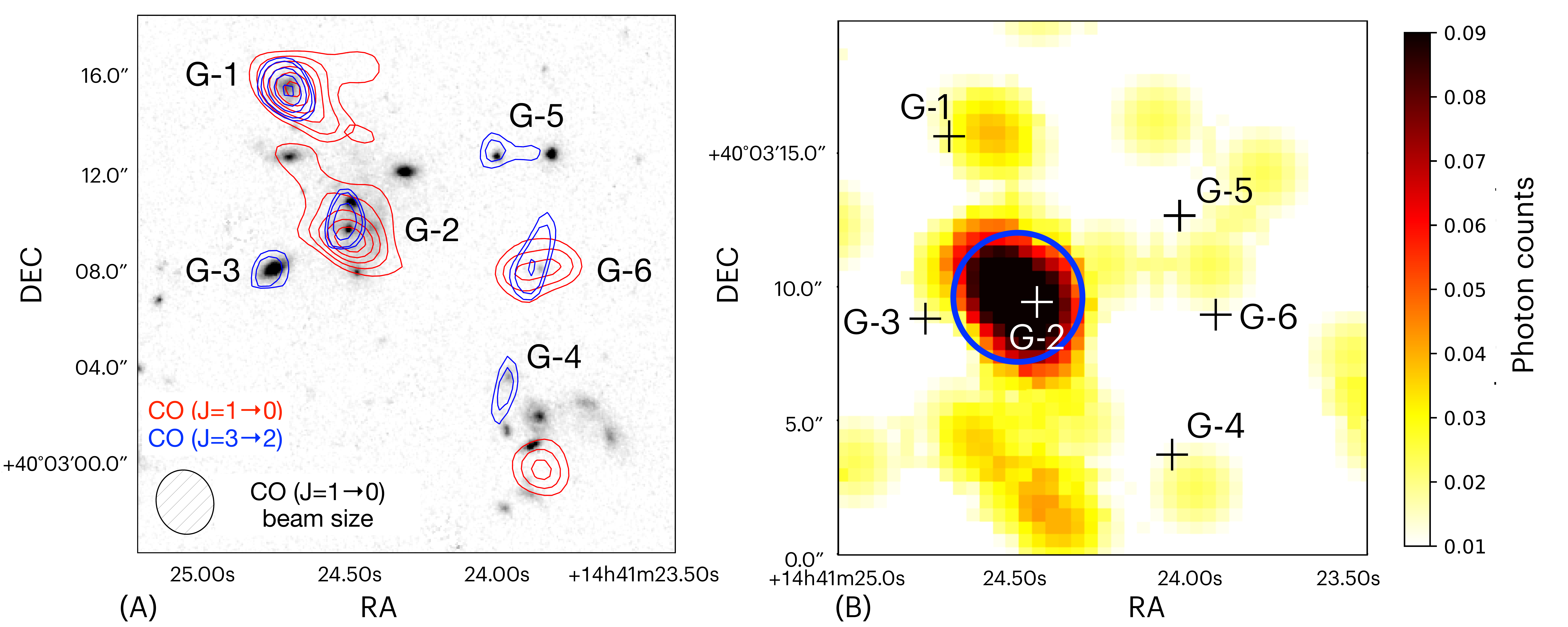}
\end{figure}
\noindent {\bf Figure S1: Images of CO (J=1→0), CO (J=3→2), and X-ray.} 
{\bf (A):} The CO (J=1→0) (red) and CO (J=3→2) (blue) emissions overlaid on the Hubble Space Telescope Image. 
The contour levels denote [2.5, 3.5, 4.5, 5.5, 6.5, 7.5]$\sigma$. 
The shaded aperture is the beam size of observations for CO (J=1→0). 
Sources with CO emissions are marked as G-1 to G-6. {\bf (B):} 
The smoothed Chandra/ACIS image of the MAMMOTH-1 field in the 2-8 keV band. 
We smooth the image with a Gaussian kernel of 5 pixels \cite{Gultekin2021}.
The blue circle shows the aperture used to extract the flux with the radius of $3''$. 
Galaxies at $z\approx2.31$ are marked as G-1 to G-6.

\newpage
\begin{figure}[b!t]
\centering
\includegraphics[width=\textwidth]{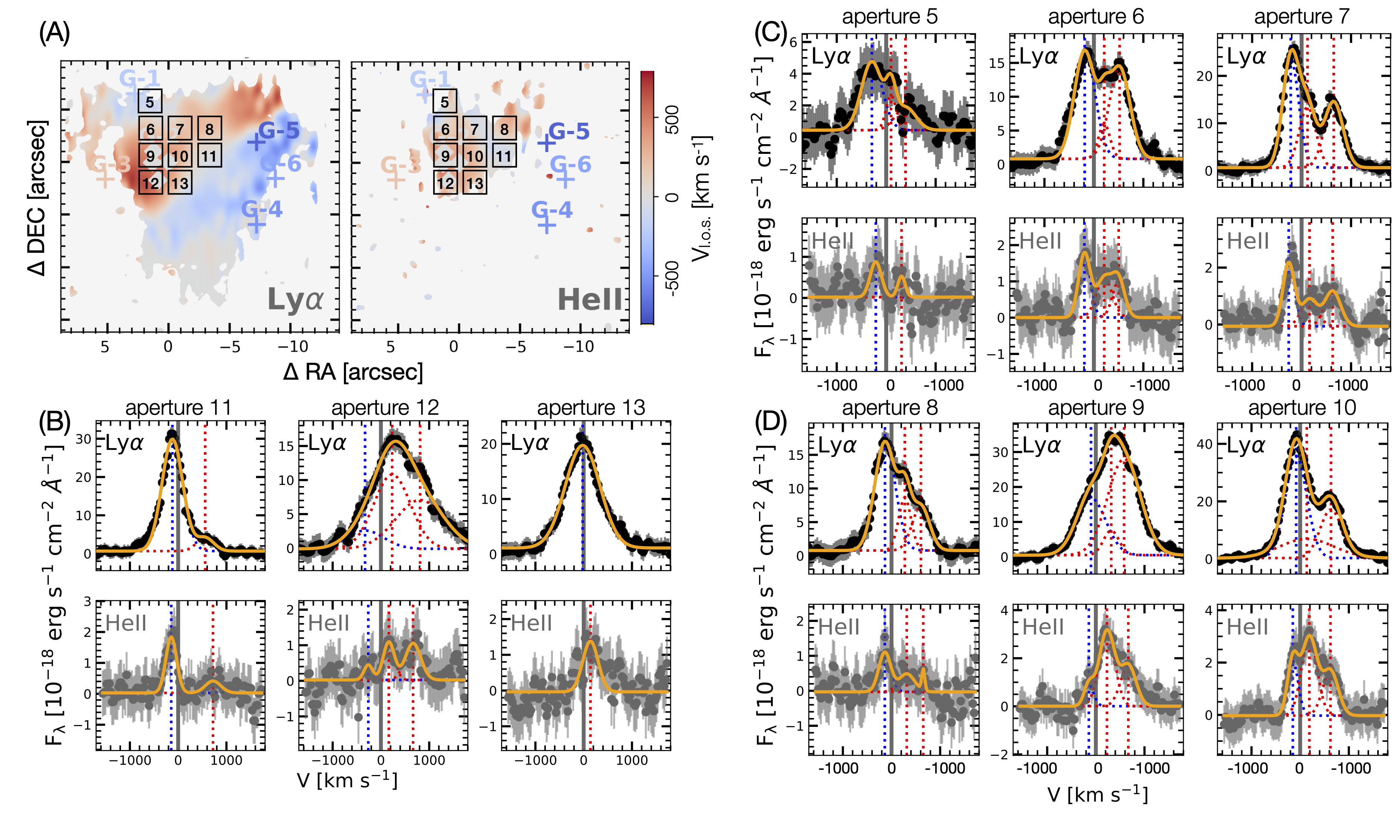}
\end{figure}
\noindent {\bf Figure S2: Spectra of Ly$\alpha$ and \heii.}
{\bf (A):} The arrangement of the nine $2.3''\times2.3''$ apertures from which the spectra are extracted.
{\bf (B-D):} the dots are the data points and the orange lines are results from the multiple Gaussian fitting.
The error bar represents the 1-$\sigma$ noise which indicates the 68\% confidence range.
The individual Gaussians are shown by the dashed lines.
The Ly$\alpha$ line has a similar Gaussian fit with that of the \heii. 
Because \heii \ is a non-resonant line, the multiple Gaussians are mainly caused by the gas motion in the circumgalactic 
medium (CGM).




\newpage
\begin{figure}[b!t]
\centering
\includegraphics[width=\textwidth]{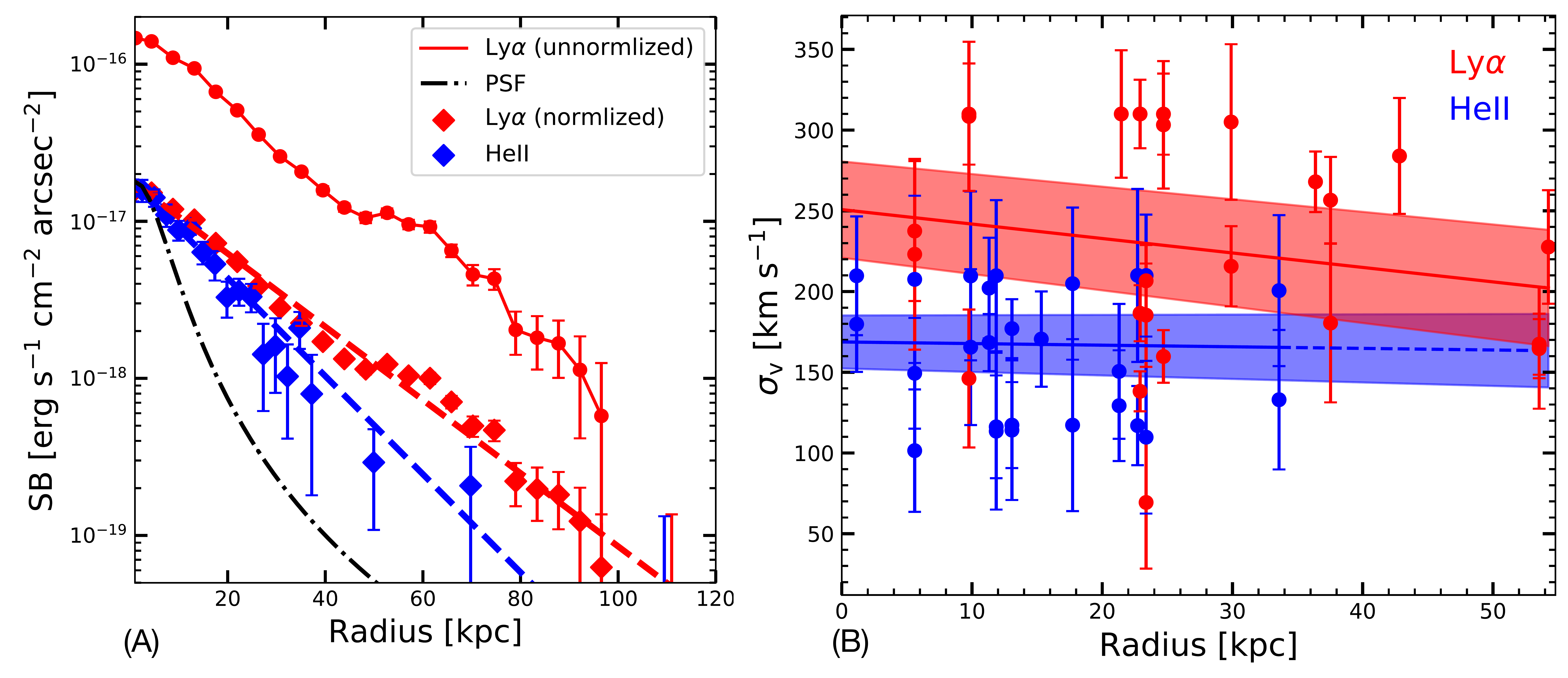}
\end{figure}
\noindent {\bf  Figure S3: SB profile and velocity dispersion profile of Ly$\alpha$ and \heii.}
{\bf (A):}
The SB profile of the Ly$\alpha$ (red) and the \heii \ (blue) emission.
For the Ly$\alpha$, the red solid line represents the true SB profile, while the red diamonds are the SB profile with the peak flux normalized to the peak flux of \heii. 
The dashed lines are the results of fitting with Eq.~\ref{SB_profile_fit}.
Under the same flux level, the Ly$\alpha$ shows a very similar SB profile to \heii. This indicates that RT effects are not the dominant powering mechanism.
{\bf (B):} The velocity dispersion profile of Ly$\alpha$ (red points) and \heii \ (blue points).
The solid lines are linear models fitted to the data. 
Ly$\alpha$ has a decreasing profile; while \heii \ has a constant profile. 
This indicates that the RT effect becomes weaker further away from the inner region. 
The error bars of these two figures show the 1-$\sigma$ scatter which indicates the 68\% confidence range.


\newpage
\begin{figure}[b!t]
\centering
\includegraphics[width=\textwidth]{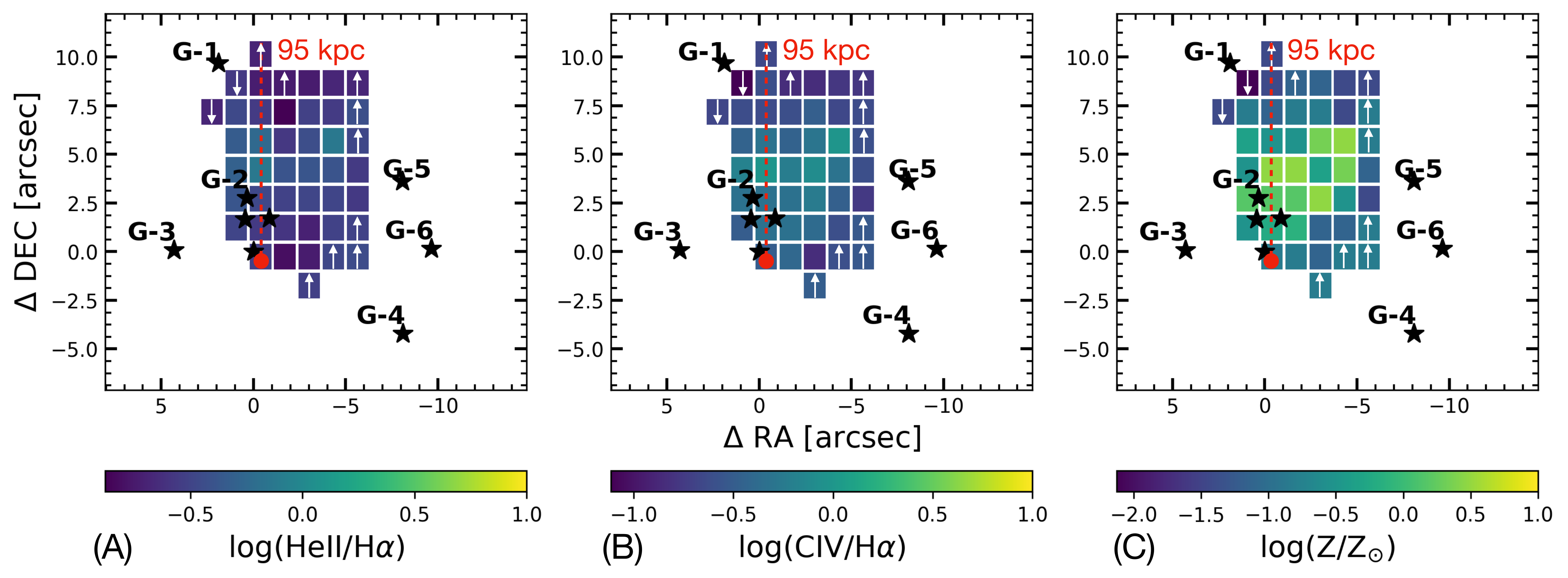}
\end{figure}
\noindent  {\bf Figure S4: Maps of line ratios and metallicity.} 
{\bf (A):} The map of log(HeII/H$\alpha$). 
Black stars mark the positions of galaxies as Fig.~1 shows. 
The south-east area, marked as a red point, is selected as the origin to measure the distance (grayscale shown in Fig.~3.
The red dashed line marks the physical scale of 95 kpc.
The up (down) white arrow within the aperture shows the value extracted from the aperture is lower (upper) limit.
We use $1.5''\times1.5''$ apertures to extract the line ratios, corresponding to the line ratio measurements shown in Fig.~3.
Colors encode the corresponding value. 
{\bf (B):} The map of log(CIV/H$\alpha$).
{\bf (C):} The map of log($Z/Z_{\odot}$).


\newpage
\begin{figure}[b!t]
\centering
\includegraphics[width=\textwidth]{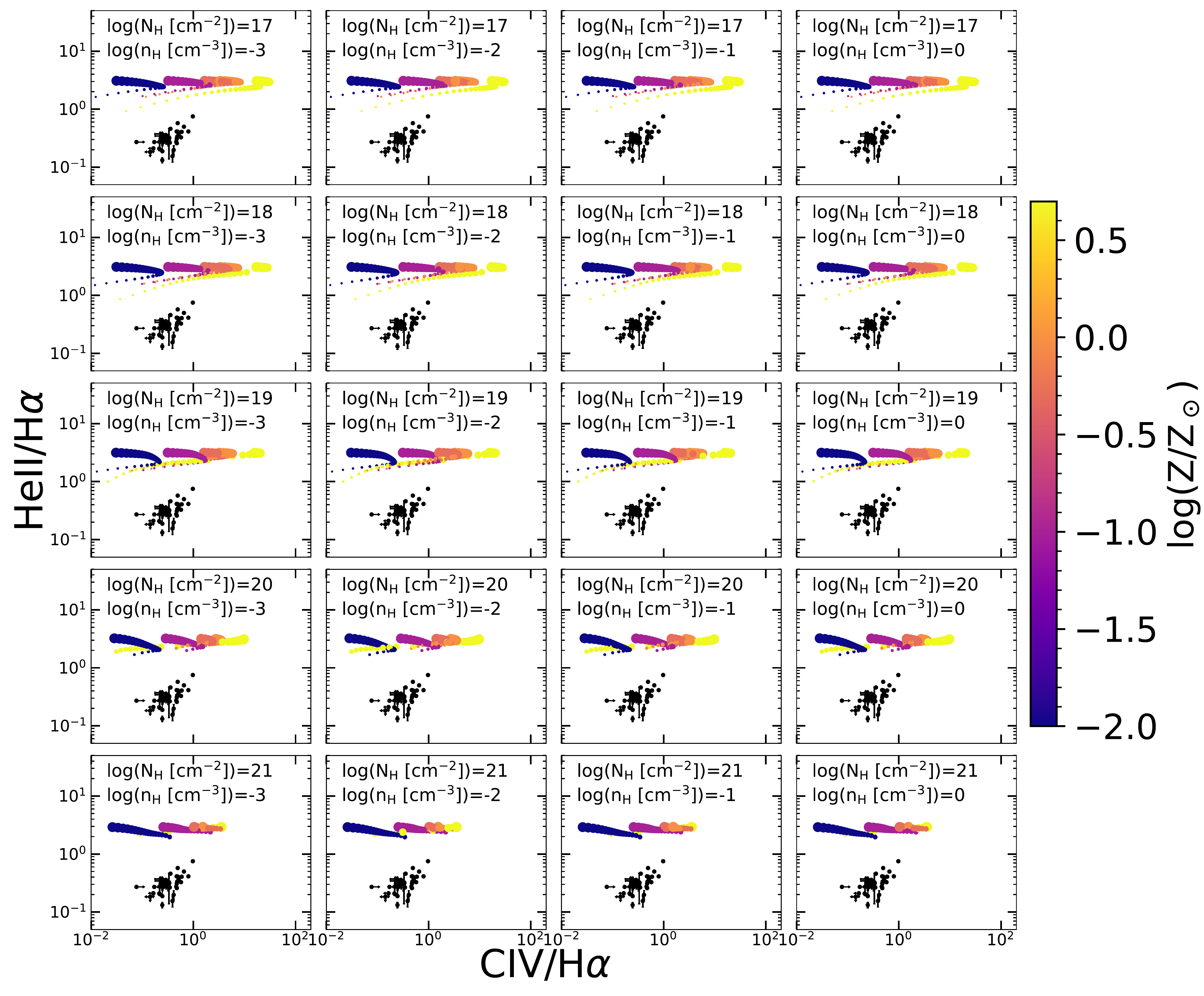}
\end{figure}
\noindent {\bf  Figure S5: Line ratios for photoionization scenario.}
The black dots are measurements the observations and the colored dots are the predictions of the photoionization model \cite{Ferland2017}, where color represents metallicities. 
The different sizes of the color dots denote different ionization parameters $\log U$. 
The largest dot denotes $\log U=0$ and the smallest dot denotes $\log U=-3$. 
The error bars of the black dots show the 2-$\sigma$ scatter which indicates the 95\% confidence range.
Over the calculated parameter space \cite{Arrigoni2015a, Cai2017a}, 
the observed line ratio of \heii /H$\alpha$ is one order of magnitude smaller than the 
predicted values from the photoionization model.

\newpage
\begin{figure}[b!t]
\centering
\includegraphics[width=\textwidth]{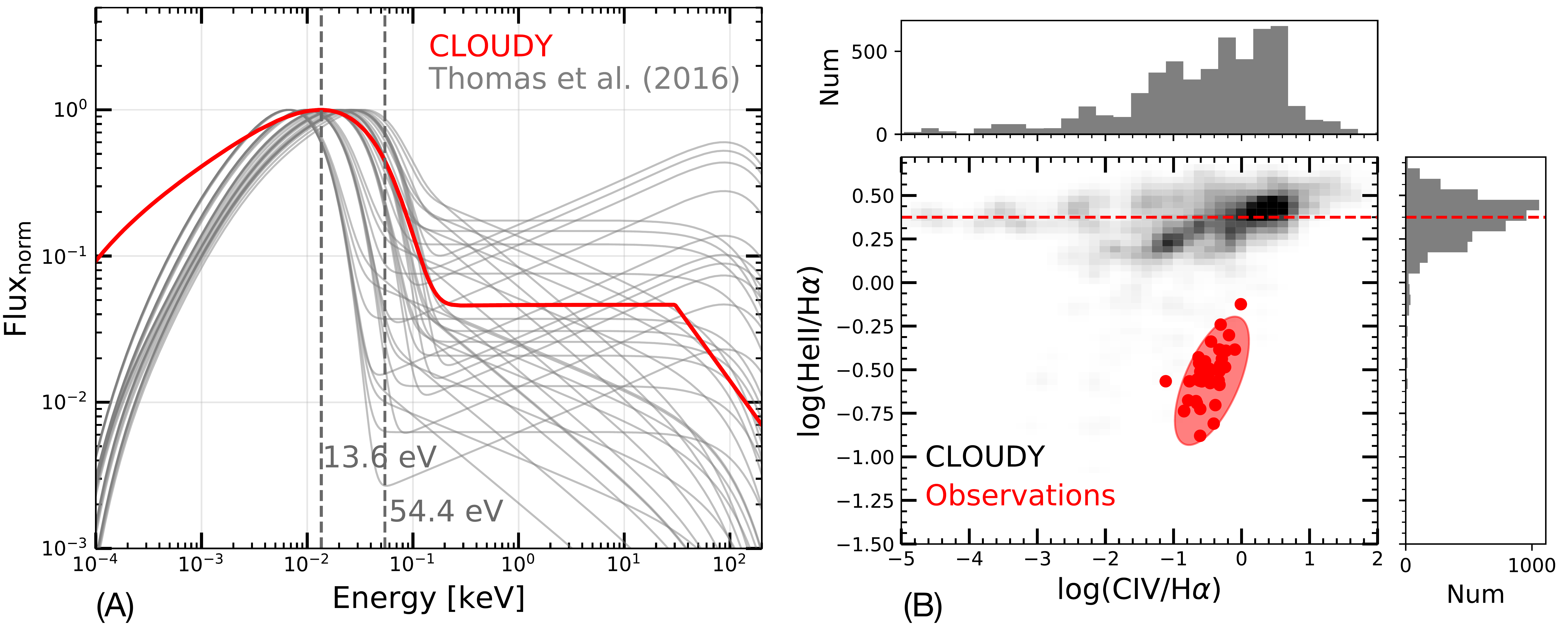}
\end{figure}
\noindent {\bf  Figure S6: Examination of different AGN continua.}
{\bf (A):}
The red line shows the fiducial AGN template from \textsc{CLOUDY}, and the gray lines show the diversity of possible AGN continua we considered.
The two dashed vertical lines mark the energy of hydrogen and helium ionizing potentials.
The y-axis shows the normalized flux which is normalized by the peak value.
The $E_{\rm peak}$ of the gray lines ranges from $-2\leq{\rm log}(E_{\rm peak}/{\rm keV})\leq-1.8$.
{\bf (B):}
The 2D histogram with the two marginal histograms of line ratios.
The 5400 models generated using \textsc{CLOUDY}, we measure an average line ratio of \heii/H$\alpha$=$2.3\pm0.8$, and the observed value is \heii/H$\alpha$=$0.3\pm0.1$.  
The red ellipse shows the 2-$\sigma$ uncertainty of the observed line ratios.
Only 17 models ($\sim0.3$\%) fall into the red cloud among the 5400 models. 
This further indicates that the pure photoionization model cannot interpret observations.



\newpage
\begin{figure}[b!t]
\centering
\includegraphics[width=\textwidth]{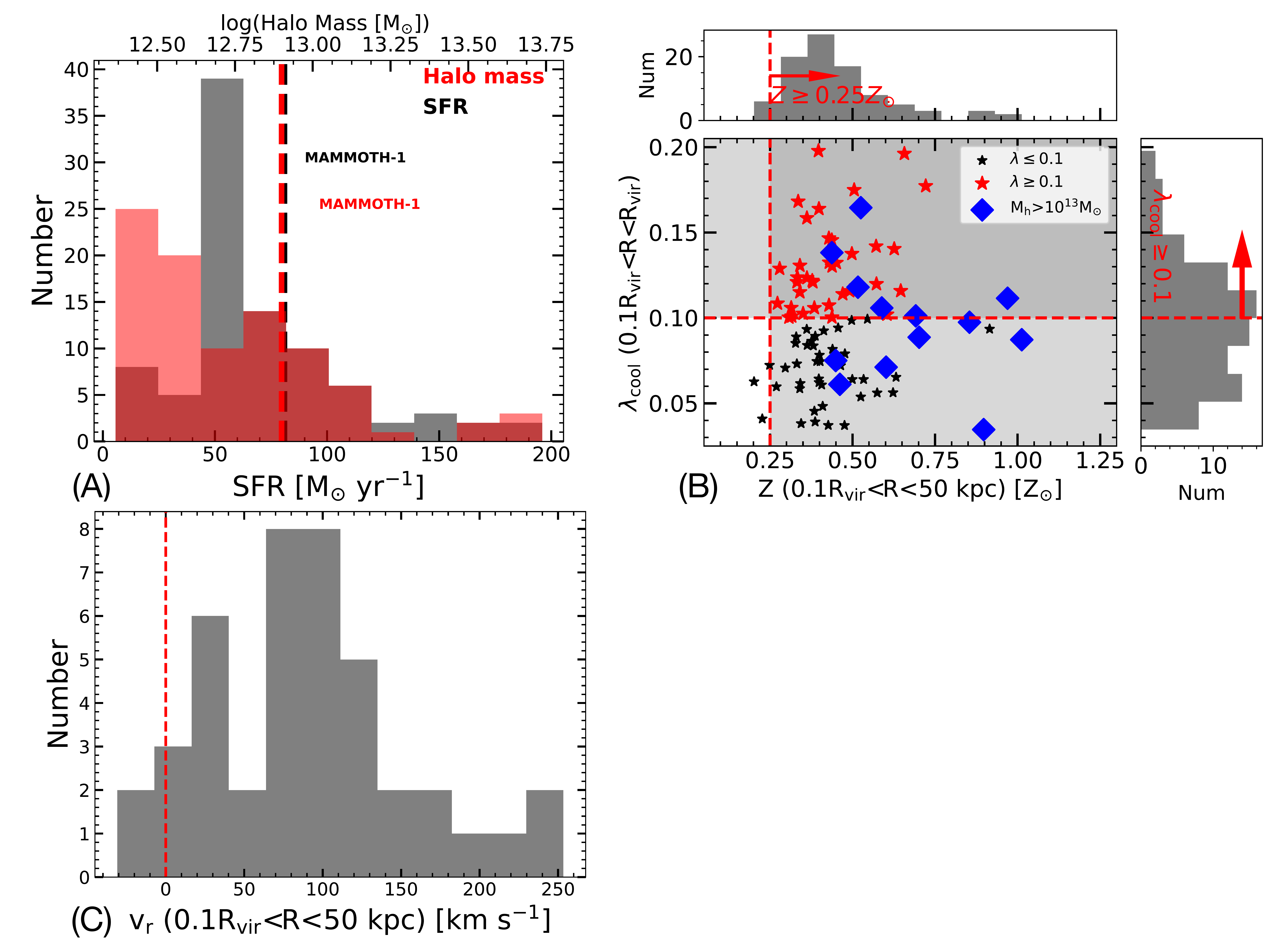}
\end{figure}
\noindent {\bf  Figure S7: Properties of simulated systems at $z=2$ compared to MAMMOTH-1.} {\bf (A):} Properties of galaxies in galaxy groups selected from TNG-100. 
The red and black histograms show the distribution of halo masses and star formation rates, respectively. 
The two dashed lines mark the corresponding values for G-2. 
{\bf (B):} The spin parameter versus the metallicity within 50 kpc of 91 MAMMOTH-like systems. 
The blue diamonds represent the systems with the host halo mass of $M_{\rm h}\geq 10^{13} \ M_{\odot}$ and the black stars represent the system with the host halo mass of $M_{\rm h}\leq 10^{13} \ M_{\odot}$. 
{\bf (C):} The histogram of the average radial velocity of the cool gas within the annulus of $\rm 0.1R_{\rm vir}\leq r\leq 50 \ {\rm kpc}$. 
The positive value indicates that the gas is flowing into the galaxy.



\newpage
\begin{figure}[b!t]
\centering
\includegraphics[width=\textwidth]{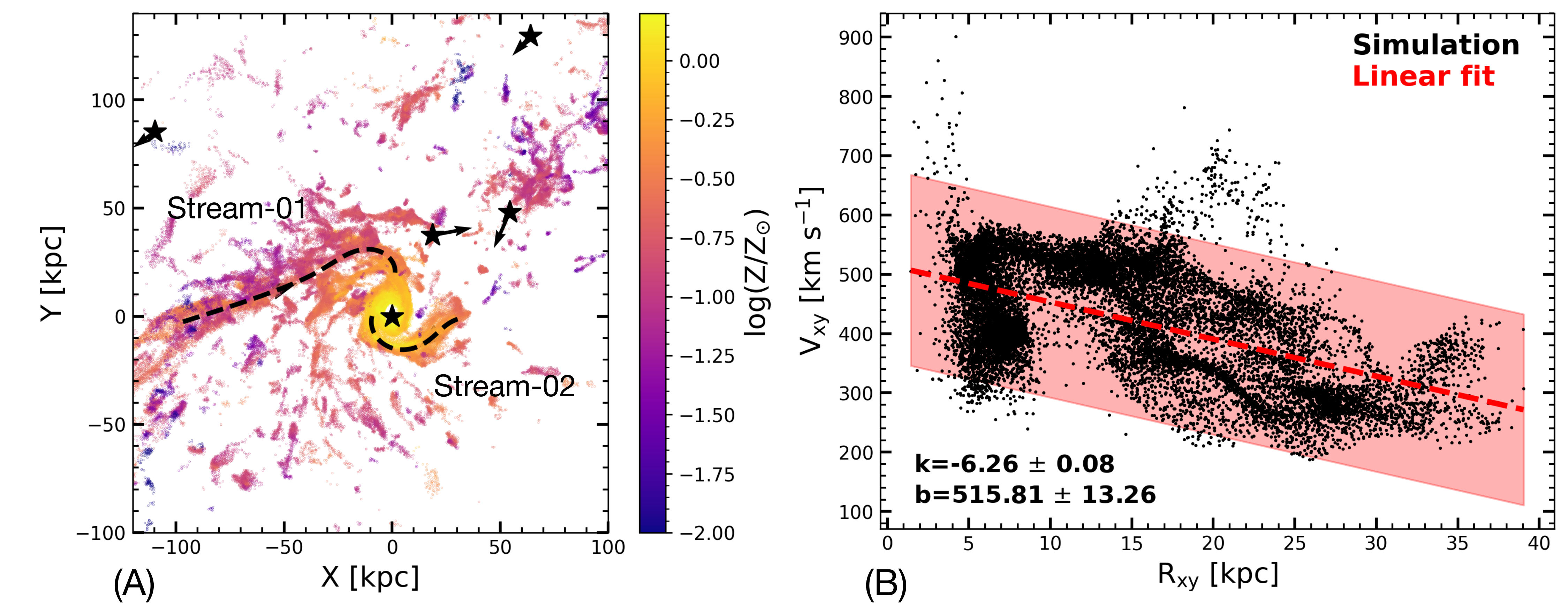}
\end{figure}
\noindent  {\bf Figure S8: Example of cool gas kinematics in simulations.} {\bf (A):} A simulated system taken from the TNG-100 box \cite{Nelson2019}, which has similar properties to MAMMOTH-1.
The color encodes the metallicity. 
The cool gas ranges from a few tenths to one solar metallicity ($Z_{\odot}$) on the scale of 100 kpc, consistent with our observations. 
Black stars mark the halos in this system. 
Two major inspiraling streams were fitted using Eq.~\ref{spiral_func} (black dashed lines). 
{\bf (B):} The velocity profile of stream-02 in panel A. 
Black dots represent the gas particles from simulations, the red dashed line is a linear model fitted to the points, and the red box denotes the 2-$\sigma$ scatter.

\newpage
\begin{figure}[b!t]
\centering
\includegraphics[width=0.82\textwidth]{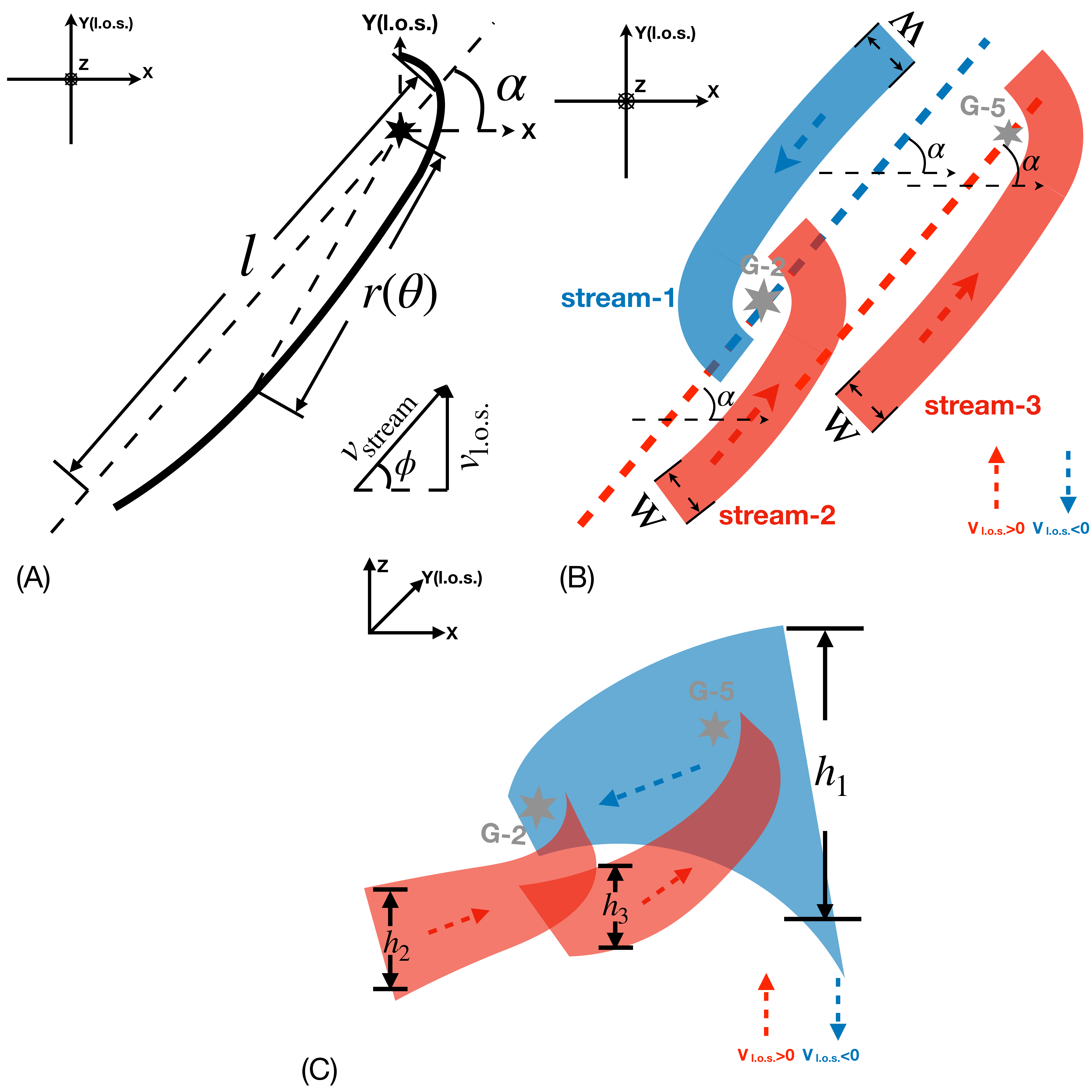}
\end{figure}
\noindent {\bf Figure S9: Geometry of the inspiraling-stream model.} {\bf (A):} The geometry of a single inspiraling stream projected on the $x-y$ plane, where $l$ is the length of the 
stream, $r$ is the radius to the central source of the stream, $\alpha$ is the angle between the stream and the $x$-axis, and $\phi$ is the angle between the gas particle velocities and the direction of the line of sight. 
The $y$-axis is fixed as the sightline of observations. 
{\bf (B):} The geometric arrangement of the three streams. Two of them are associated with G-2 and the third is associated with G-5. 
The width, $W$, and the angle between the stream and $x$-axis, $\alpha$, are the same for all three streams. 
{\bf (C):} Another projection of the model.
Here, $h_{\rm{i}}$ represents the thickness of the 
three streams. 
All of the labelled parameters are fixed based on either simulations or observations.





\newpage
\begin{figure}[b!t]
\centering
\includegraphics[width=0.9\textwidth]{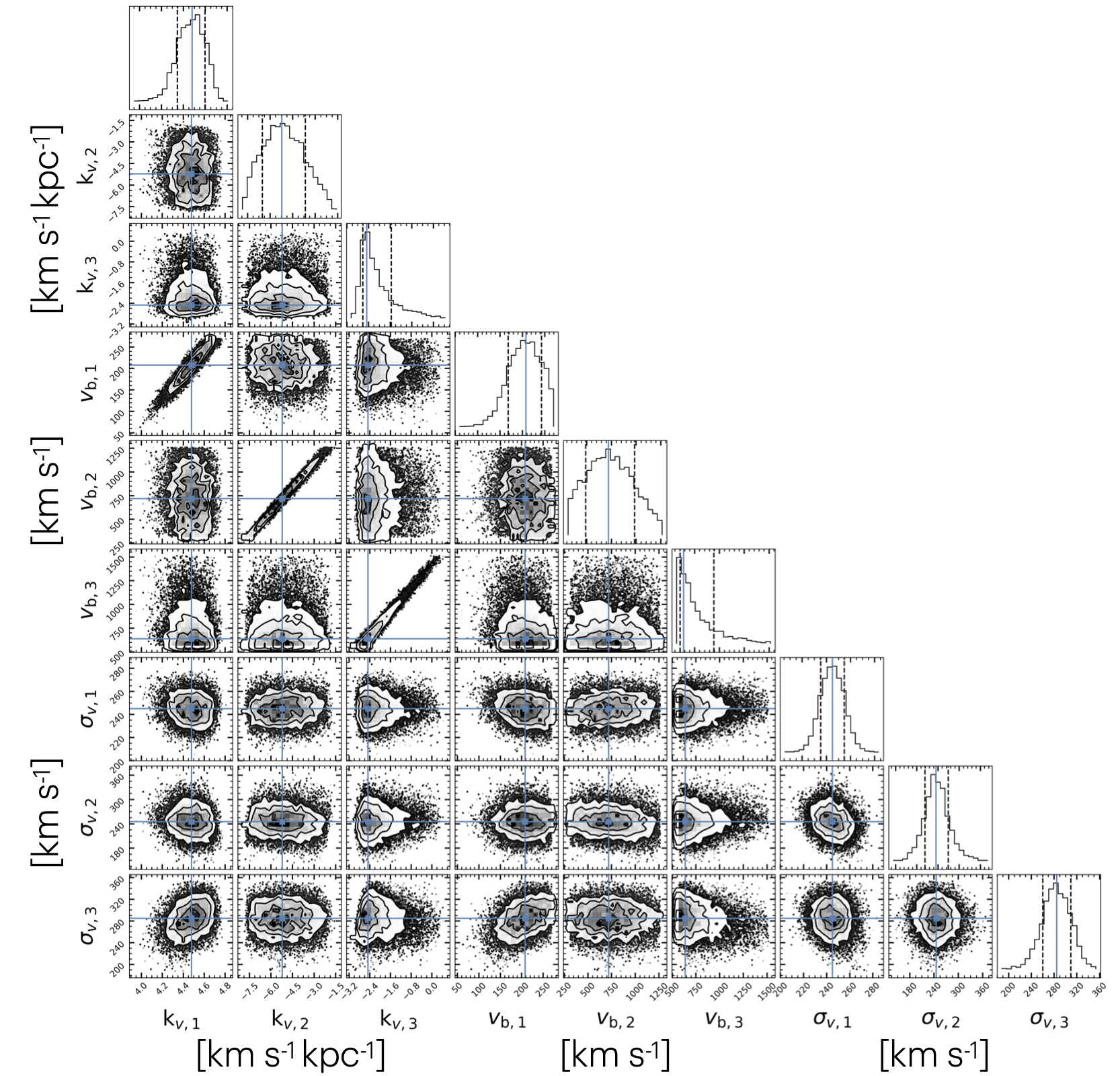}
\end{figure}
\noindent {\bf Figure S10: MCMC results.} 
Corner plot of the MCMC results used to determine the parameters of the inspiraling stream model. 
The one and two-dimensional posterior probabilities for the nine free kinematic parameters are shown. 
For $\rm k_{v,1}$, $\rm k_{v,2}$, and $\rm k_{v,3}$, the unit is $\rm km \ s^{-1} \ kpc^{-1}$. 
For $\rm v_{b,1}$, $\rm v_{b,2}$, $\rm v_{b,3}$, $\rm \sigma_{v,1}$, $\rm \sigma_{v,2}$, and $\rm \sigma_{v,3}$, the unit is $\rm km \ s^{-1}$. 
The blue solid lines indicate the best-fitting parameters while the black dashed lines show the 16\% and 84\% percentiles for each parameter. 
The contour levels of the 2D posterior probabilities are at 0.5-$\sigma$, 1-$\sigma$, 1.5-$\sigma$, and 2-$\sigma$.
Because we adopt the linear velocity profile (Eq.~\ref{vprofile}), the slopes ($k_{v,1}$, $k_{v,2}$, and $k_{v,1}$) are correlated with the intercepts ($v_{\rm b,1}$, $v_{\rm b,2}$, and $v_{\rm b,3}$). 
The best-fitting values are listed in Tab.~S\ref{R1q7_tab}.

\newpage
\begin{figure}[b!t]
\centering
\includegraphics[width=.9\textwidth]{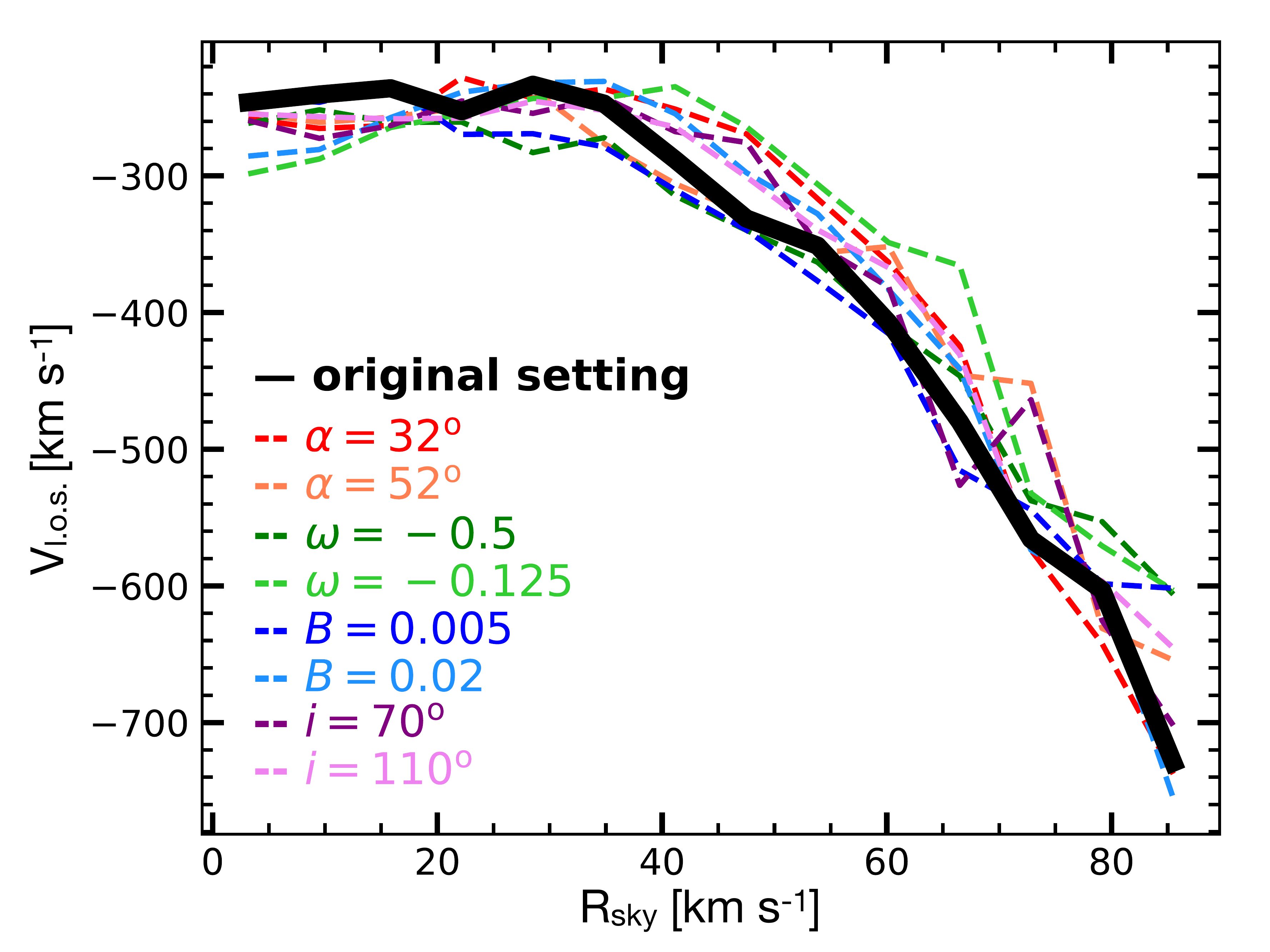}
\end{figure}
\noindent {\bf  Figure S11: Projected-line-of-sight velocity profile.}
Here, V$_{\rm l.o.s.}$ is the line-of-sight velocity and $\rm R_{sky}$ is the distance from a gas particle to G-2 in the plane of the sky (Fig.~4B).
The black solid line is from the best-fitting configuration, while the colored dashed lines are from alternative configurations. 
Within a reasonable parameter range of $32^{\rm o}\leq\alpha \leq 52^{\rm o}$, $-0.5\leq\omega \leq -0.125$, $0.005\leq B \leq 0.02$, and $70^{\rm o}\leq i \leq 110^{\rm o}$,
the model yields a consistent velocity profile regardless of the geometric configuration.


\newpage
\begin{figure}[b!t]
\centering
\includegraphics[width=.9\textwidth]{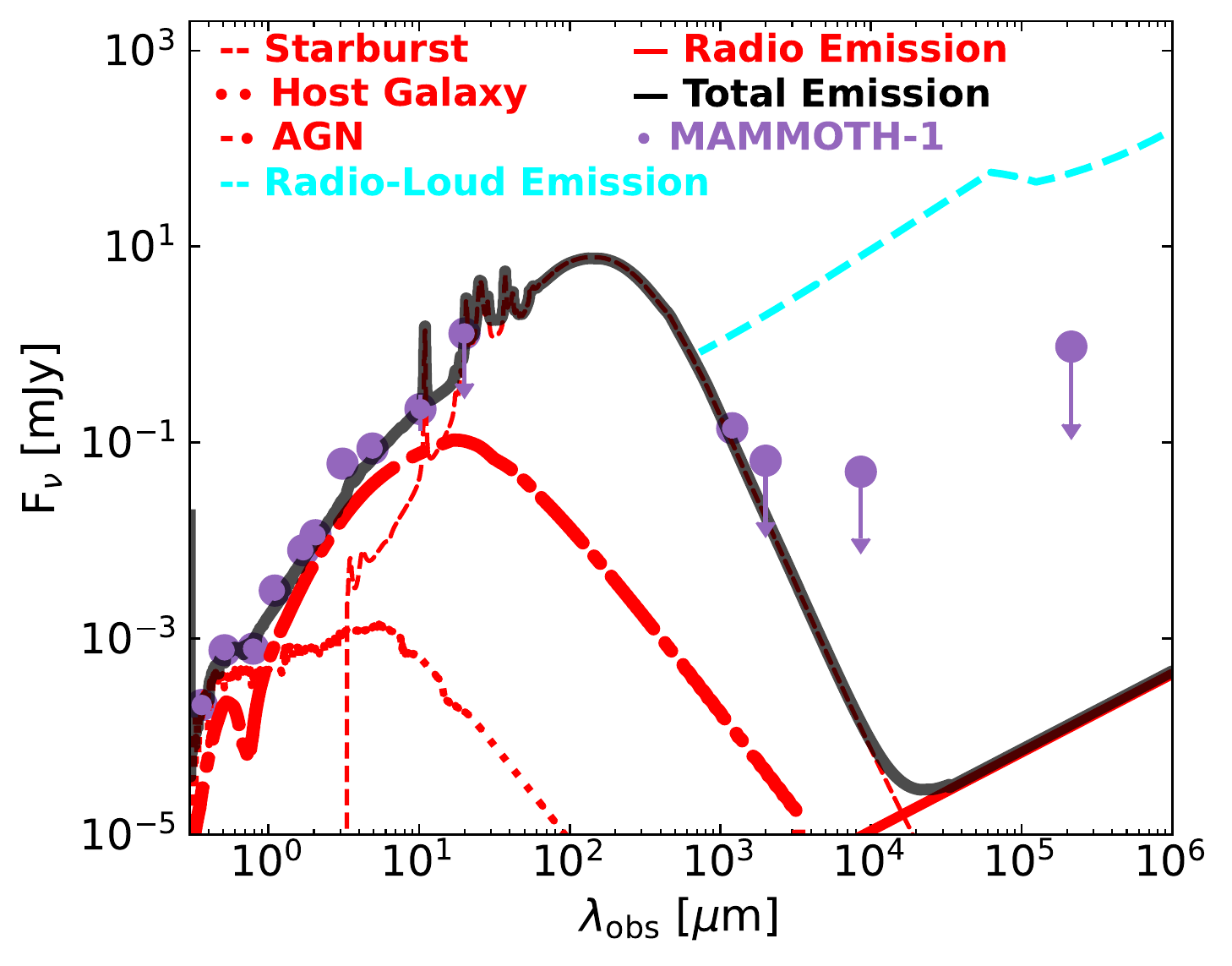}
\end{figure}
\noindent {\bf Figure S12: SED of G-2.}  
$y$-axis denotes the flux density at frequency of $\nu$ ($\rm F_{\nu}$). 
$x$-axis denotes the observed wavelength ($\lambda_{\rm obs}$) in the unit of $\mu$m.
The purple dots are observations from the ultraviolet to the radio. 
The black line is the best-fitting model SED output by \textsc{Cigale} \cite{Boquien2019}. 
The red lines with different line styles and widths are the different components. 
The cyan dashed line is the stacked SED of radio-loud quasars \cite{Shang2011}.


\newpage
\begin{figure}[b!t]
\centering
\includegraphics[width=0.6\textwidth]{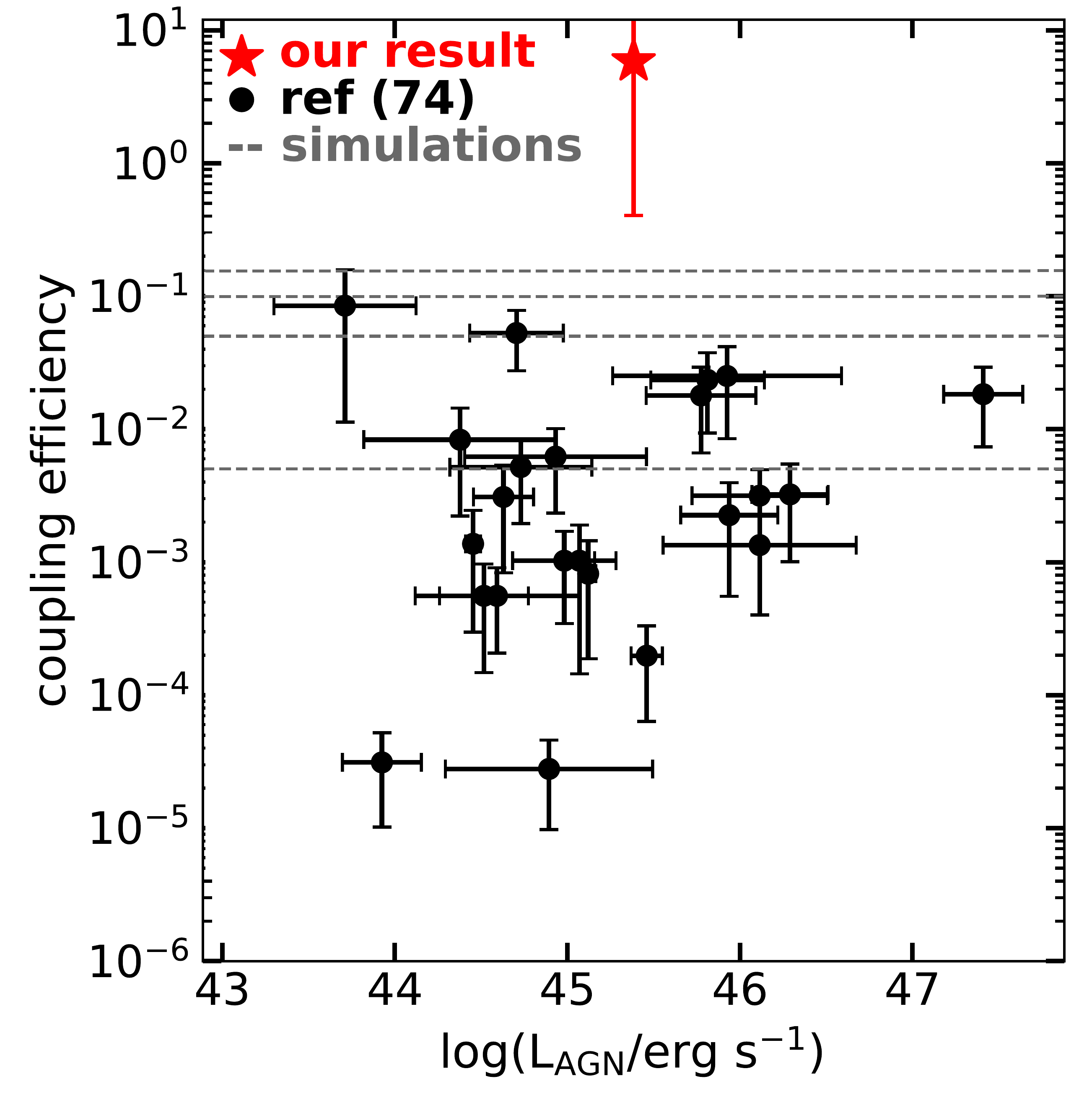}
\end{figure}
\noindent {\bf  Figure S13: Coupling efficiency.} 
A comparison of the coupling efficiency calculated from our observations, assuming an AGN outflow model (red star) and other observations (black dots) and simulations (dashed lines) \cite{Harrison2018}. 
The error bar of our result represents the upper and lower limit. 
The coupling efficiency of $f_{\rm c}=5.82$ is much higher than those from simulations and previous observations. 

\newpage
\begin{table}
\centering
\caption{ {\bf The best-fitting velocities of the multiple Gaussians in Fig.~2.}
$v_{\rm b}$ is the velocity of the blue component, $v_{\rm m}$ is the velocity of the middle component, and $v_{\rm r}$ is the velocity of the red component. 
For spectra extracted from Aperture 4, two Gaussians are sufficient to provide good fitting. 
Thus, only two velocity values are provided.
The velocities of multiple components of Ly$\alpha$ are within of the values of \heii.
This indicates that Ly$\alpha$ traces the cool gas kinematics.} 
\begin{tabular}[0.8\textwidth]{lllllll} 
\hline
\hline
 \multicolumn{1}{l|}{}                             & \multicolumn{3}{c|}{Ly$\alpha$}                                                         & \multicolumn{3}{c}{\heii}                                                         \\ 
\cline{2-7}
\multicolumn{1}{c|}{}         & \multicolumn{1}{c}{$v_{\rm b}$ [km s$^{-1}$]} & \multicolumn{1}{c}{$v_{\rm m}$ [km s$^{-1}$]} & \multicolumn{1}{c|}{$v_{\rm r}$ [km s$^{-1}$]} & \multicolumn{1}{c}{$v_{\rm b}$ [km s$^{-1}$]} & \multicolumn{1}{c}{$v_{\rm m}$ [km s$^{-1}$]} & \multicolumn{1}{c}{$v_{\rm r}$ [km s$^{-1}$]}  \\ 
\cline{2-7}
\multicolumn{1}{c|}{Aperture 1} & \multicolumn{1}{c}{$-169\pm35$} & \multicolumn{1}{c}{$196\pm53$}    & \multicolumn{1}{c|}{$535\pm43$} & \multicolumn{1}{c}{$-198\pm39$} & \multicolumn{1}{c}{$231\pm62$}    & \multicolumn{1}{c}{$527\pm24$}  \\
\multicolumn{1}{c|}{Aperture 2} & \multicolumn{1}{c}{$-138\pm41$} & \multicolumn{1}{c}{$228\pm46$}    & \multicolumn{1}{c|}{$594\pm57$} & \multicolumn{1}{c}{$-186\pm38$} & \multicolumn{1}{c}{$261\pm39$}    & \multicolumn{1}{c}{$718\pm67$}  \\
\multicolumn{1}{c|}{Aperture 3} & \multicolumn{1}{c}{$-139\pm34$} & \multicolumn{1}{c}{$363\pm48$}    & \multicolumn{1}{c|}{$813\pm55$} & \multicolumn{1}{c}{$-46\pm26$} & \multicolumn{1}{c}{$256\pm37$}    & \multicolumn{1}{c}{$680\pm42$}  \\
\multicolumn{1}{c|}{Aperture 4} & \multicolumn{1}{c}{$-103\pm23$} & \multicolumn{1}{c}{}       & \multicolumn{1}{c|}{$626\pm51$} & \multicolumn{1}{c}{$-108\pm37$} & \multicolumn{1}{c}{}       & \multicolumn{1}{c}{$613\pm25$}  \\ 
\hline    
\hline
\end{tabular}
\label{multiple_component}
\end{table}

\begin{table}
\centering
\caption{{\bf The best-fitting velocities of the multiple Gaussians in Fig.~S2.}
Same as Table S1, but for the smaller apertures shown in Fig.~S2.}
\begin{tabular}{c|ccc|ccc} 
\hline
\hline
           & \multicolumn{3}{c|}{Ly$\alpha$} & \multicolumn{3}{c}{\heii}  \\ 
\cline{2-7}
           & $v_{\rm b}$ [km s$^{-1}$] & $v_{\rm m}$ [km s$^{-1}$]& $v_{\rm r}$ [km s$^{-1}$]     & $v_{\rm b}$ [km s$^{-1}$]& $v_{\rm m}$ [km s$^{-1}$]& $v_{\rm r}$ [km s$^{-1}$]      \\ 
\cline{2-7}
Aperture 5 & $-291\pm34$ & $95\pm25$   & $389\pm41$       & $-195\pm44$ &      & $337\pm51$         \\
Aperture 6 & $-213\pm47$ & $210\pm32$  & $517\pm29$       & $-194\pm28$ & $205\pm43$  & $501\pm61$         \\
Aperture 7 & $-220\pm45$ & $119\pm30$  & $651\pm53$       & $-254\pm36$ & $151\pm31$  & $629\pm58$         \\
Aperture 8 & $-134\pm42$ & $273\pm44$  & $604\pm41$       & $-139\pm54$ & $316\pm33$  & $653\pm29$         \\
Aperture 9 & $-93\pm37$  & $307\pm41$  & $581\pm56$       & $-147\pm33$ & $223\pm28$  & $641\pm41$         \\
Aperture 10 & $-105\pm34$ & $193\pm23$  & $616\pm52$       & $-151\pm27$ & $187\pm41$  & $621\pm31$         \\
Aperture 11 & $-119\pm46$ &      & $546\pm31$       & $-135\pm25$ &      & $772\pm45$         \\
Aperture 12 & $-206\pm34$ & $230\pm48$  & $702\pm51$       & $-262\pm36$ & $163\pm23$  & $667\pm41$         \\
Aperture 13 & $-22\pm33$  &      &        &  $135\pm51$    &      &             \\
\hline
\hline
\end{tabular}
\label{R2q6_tab}
\end{table}

\begin{table}
\centering
\caption{{\bf Measurements from the CO (J=3→2) data.} 
Redshift is the spectroscopic redshifts from the CO (J=3→2). 
FWHM denotes the line width of the CO (J=3→2) emission yielded by the Gaussian fit. 
$L$ is the CO (J=3→2) luminosity \cite{Li2021}. 
$v$ is the velocity converted from the Redshift. 
The systemic redshift is adopted as $z=2.3116\pm0.0004$ of G-2 from the CO (J=1→0) observations \cite{Emonts2019}. 
The H$_{\rm 2}$ gas mass converted from the CO (J=3→2) emissions \cite{Li2021}.}
\begin{tabular}[0.8\textwidth]{c|ccccc} 
\hline
\hline
\multirow{1}{*}{Source} & \multicolumn{5}{c}{CO (J=3→2)}  \\ 
\cline{2-6}
                        & Redshift & $\rm FWHM$ [km s$^{-1}$] & $L$ [$10^{10}$ K km s$^{-1}$ pc$^{-2}$]    & $v$ [km s$^{-1}$]    &  $M_{\rm H2}$ [$10^{10} \ M_{\odot}$]   \\ 
\hline
G-1 & $2.3088\pm 0.0004$   & $184\pm32$  & $7.1\pm1.1$  & $-253\pm51$ &  $4.3\pm0.1$      \\
G-2 & $2.3120\pm0.0006$    & $370\pm91$  & $6.7\pm1.4$   & $0\pm51$ &  $4.0\pm0.1$    \\
G-3 & $2.3137\pm0.0004$   & $181\pm75$  & $3.0\pm1.1$   & $190\pm51$ &   $2.1\pm0.8$      \\
G-4 & $2.3059\pm0.0004$   & $161\pm47$  & $3.7\pm0.9$   & $-516\pm51$ &  $2.6\pm0.6$      \\
G-5 & $2.3037\pm0.0003$   & $81\pm28$   & $2.1\pm0.6$   & $-715\pm45$ &  $1.5\pm0.4$     \\
G-6 & $2.3067\pm0.0005$   & $281\pm74$  & $5.4\pm1.2$    & $-443\pm57$ & $3.6\pm0.1$     \\
\hline
\hline
\end{tabular}
\end{table}


\begin{table}
\centering
\caption{{\bf The parameters used to construct the inspiraling stream model.}
The first column lists the parameters, and the following three columns list the values used for the three streams.
The last column gives a brief description of the constraints on the parameters.}
\begin{tabular}{p{0.2\linewidth}|p{0.18\linewidth}p{0.18\linewidth}p{0.18\linewidth}p{0.2\linewidth}} 
\hline
\hline
               & \textbf{Stream-1} & \textbf{Stream-2} & \textbf{Stream-3} & {\bf Constraints}                                                           \\ 
\hline
\hline
\textbf{A [kpc]}     & 177               & 177               & 177               & {\footnotesize Determined from simulations.}  \\ 
\textbf{B}     & 0.01             & 0.01             & 0.01             & {\footnotesize Determined from simulations.}                               \\
\textbf{$\omega$} & -0.25            & -0.25            & -0.25            & {\footnotesize Determined from simulations.}                               \\
\textbf{$\alpha$ [deg]} & 222              & 42               & 42               & {\footnotesize Determined from observations.}  \\
\textbf{$\theta$ [rad]} &  [$1.24\pi$, $2.23\pi$]                &   [$0.27\pi$, $1.23\pi$]               &       [$0.25\pi$, $1.23\pi$]           & {\footnotesize Determined from observations.}                                                                    \\
\textbf{h [kpc]}     & 120              & 55               & 55               & {\footnotesize Determined from observations.}            \\
\textbf{$i$ [deg]}     & 90               & 90               & 90               & {\footnotesize Assumption.}                                                 \\
\textbf{W [kpc]}     & 55               & 55               & 55               & {\footnotesize Assumption.}                \\
\hline
$k_{v}$  [km s$^{-1}$ kpc$^{-1}$]   & $4.5^{+0.1}_{-0.1}$               & $-5.2^{+2.0}_{-1.4}$               & $-2.3^{+0.7}_{-0.3}$               & {\footnotesize determined from MCMC.}                \\
$v_{b}$  [km s$^{-1}$]   &   $210^{+36}_{-38}$              & $730^{+256}_{-237}$                & $640^{+242}_{-100}$               & {\footnotesize determined from MCMC.}                \\
$\sigma_{v}$  [km s$^{-1}$]   & $250^{+10}_{-10}$                & $250^{+28}_{-27}$               & $280^{+25}_{-24}$               & {\footnotesize determined from MCMC.}                \\
\hline
\hline
\end{tabular}
\label{R1q7_tab}
\end{table}


\begin{thebibliography}{10}
\expandafter\ifx\csname url\endcsname\relax
  \def\url#1{\texttt{#1}}\fi
\expandafter\ifx\csname urlprefix\endcsname\relax\def\urlprefix{URL }\fi
\providecommand{\bibinfo}[2]{#2}
\providecommand{\eprint}[2][]{\url{#2}}

\bibitem{Tumlinson2017}
J.~{Tumlinson}, M.~S. {Peeples}, J.~K. {Werk}, {\it \araa\/} {\bf 55}, 389  (2017).
  
\bibitem{Keres2005}
D.~{Kere{\v{s}}}, N.~{Katz}, D.~H. {Weinberg}, R.~{Dav{\'e}}, {\it \mnras\/}
  {\bf 363}, 2 (2005).  

\bibitem{Stern2020}
J.~{Stern}, D.~{Fielding}, C.-A. {Faucher-Gigu{\`e}re}, E.~{Quataert}, {\it  \mnras\/} {\bf 492}, 6042 (2020).  

\bibitem{Stewart2017}
K.~R. {Stewart}, {\it et~al.\/}, {\it \apj\/} {\bf 843}, 47 (2017).

\bibitem{Suresh2019}
J.~{Suresh}, D.~{Nelson}, S.~{Genel}, K.~H.~R. {Rubin}, L.~{Hernquist}, {\it
  \mnras\/} {\bf 483}, 4040 (2019).
  
\bibitem{Lehner2019}
N.~{Lehner}, {\it et~al.\/}, {\it \apj\/} {\bf 887}, 5 (2019).
  
\bibitem{Angles-Alcazar2017}
D.~{Angl{\'e}s-Alc{\'a}zar}, {\it et~al.\/}, {\it \mnras\/} {\bf 470}, 4698  (2017).

\bibitem{Grand2019}
R.~J.~J. {Grand}, {\it et~al.\/}, {\it \mnras\/} {\bf 490}, 4786 (2019).

\bibitem{Oppenheimer2010}
B.~D. {Oppenheimer}, {\it et~al.\/}, {\it \mnras\/} {\bf 406}, 2325 (2010).

\bibitem{Brennan2018}
R.~{Brennan}, {\it et~al.\/}, {\it \apj\/} {\bf 860}, 14 (2018).

\bibitem{Prochaska2014}
J.~X. {Prochaska}, M.~W. {Lau}, J.~F. {Hennawi}, {\it \apj\/} {\bf 796}, 140  (2014).

\bibitem{Cai2017a}
Z.~{Cai}, {\it et~al.\/}, {\it \apj\/} {\bf 837}, 71 (2017).  


\bibitem{Arrigoni2018a}
\bibinfo{author}{F. Arrigoni Battaia,} \emph{et~al.},
\newblock \emph{\bibinfo{journal}{Astron. Astrophys.}} \textbf{\bibinfo{volume}{620}},
  \bibinfo{pages}{A202} (\bibinfo{year}{2018}).

\bibitem{Emonts2019}
B.~H.~C. {Emonts}, Z.~{Cai}, J.~X. {Prochaska}, Q.~{Li}, M.~D. {Lehnert}, {\it  \apj\/} {\bf 887}, 86 (2019).

\bibitem{Methods}
Materials and methods are available as supplementary materials.

\bibitem{Li2021}
Q.~{Li}, {\it et~al.\/}, {\it \apj\/} {\bf 922}, 236 (2021).

\bibitem{Yang2014}
Y.~{Yang}, A.~{Zabludoff}, K.~{Jahnke}, R.~{Dav{\'e}}, {\it \apj\/} {\bf 793},  114 (2014).

\bibitem{Arrigoni2015a}
F.~{Arrigoni Battaia}, {\it et~al.\/}, {\it \apj\/} {\bf 804}, 26 (2015).

\bibitem{Allen2008}
M.~G. {Allen}, B.~A. {Groves}, M.~A. {Dopita}, R.~S. {Sutherland}, L.~J.  {Kewley}, {\it \apjs\/} {\bf 178}, 20 (2008).

\bibitem{Steidel2014}
C.~C. {Steidel}, {\it et~al.\/}, {\it \apj\/} {\bf 795}, 165 (2014).

\bibitem{Tumlinson2011}
J.~{Tumlinson}, {\it et~al.\/}, {\it Science\/} {\bf 334}, 948 (2011).

\bibitem{Wotta2019}
C.~B. {Wotta}, {\it et~al.\/}, {\it \apj\/} {\bf 872}, 81 (2019).

\bibitem{Steidel2010}
C.~C. {Steidel}, {\it et~al.\/}, {\it \apj\/} {\bf 717}, 289 (2010).

\bibitem{Lehner2016}
N.~{Lehner}, J.~M. {O'Meara}, J.~C. {Howk}, J.~X. {Prochaska}, M.~{Fumagalli},  {\it \apj\/} {\bf 833}, 283 (2016).

\bibitem{Fabian2012}
A.~C. {Fabian}, {\it \araa\/} {\bf 50}, 455 (2012).

\bibitem{Richings2018}
A.~J. {Richings}, C.-A. {Faucher-Gigu{\`e}re}, {\it \mnras\/} {\bf 478}, 3100  (2018).

\bibitem{Circosta2018}
C.~{Circosta}, {\it et~al.\/}, {\it \aap\/} {\bf 620}, A82 (2018).

\bibitem{Stewart2011}
K.~R. {Stewart}, {\it et~al.\/}, {\it \apj\/} {\bf 738}, 39 (2011).

\bibitem{Nelson2019}
D.~{Nelson}, {\it et~al.\/}, {\it Computational Astrophysics and Cosmology\/} {\bf 6}, 2 (2019).


\bibitem{Lan2019}
T.-W. {Lan}, H.~{Mo}, {\it \mnras\/} {\bf 486}, 608 (2019).

\bibitem{Afruni2019}
A.~{Afruni}, F.~{Fraternali}, G.~{Pezzulli}, {\it \aap\/} {\bf 625}, A11  (2019).


\bibitem{Morrissey2018}
P.~{Morrissey}, {\it et~al.\/}, {\it \apj\/} {\bf 864}, 93 (2018).

\bibitem{Suzuki2008}
R.~{Suzuki}, {\it et~al.\/}, {\it \pasj\/} {\bf 60}, 1347 (2008).

\bibitem{Garmire2003}
G.~{Garmire}, {\it et~al.\/}, {\it \xgtia\/} {\bf 4851}, 28 (2003).

\bibitem{CIAO2006}
A.~{Fruscione}, {\it et~al.\/}, {\it \spiecs\/} {\bf 6270}, 62701V (2006).

\bibitem{HI4PI-Collaboration2016}
{HI4PI Collaboration}, {\it et~al.\/}, {\it \aap\/} {\bf 594}, A116 (2016).

\bibitem{Gilli2007}
R.~{Gilli}, A.~{Comastri}, G.~{Hasinger}, {\it \aap\/} {\bf 463}, 79 (2007).

\bibitem{KCWI_pipeline}
{KCWI Pipeline}, {https://github.com/Keck-DataReductionPipelines/KcwiDRP}

\bibitem{Cai2019}
Z.~{Cai}, {\it et~al.\/}, {\it \apjs\/} {\bf 245}, 23 (2019).



\bibitem{Arrigoni2018b}
F.~{Arrigoni Battaia}, {\it et~al.\/}, {\it \mnras\/} {\bf 473}, 3907 (2018).

\bibitem{Borisova2016}
E.~{Borisova}, {\it et~al.\/}, {\it \apj\/} {\bf 831}, 39 (2016).



\bibitem{Cantalupo2012}
S.~{Cantalupo}, S.~J. {Lilly}, M.~G. {Haehnelt}, {\it \mnras\/} {\bf 425}, 1992  (2012).


\bibitem{Cantalupo2005}
S.~{Cantalupo}, C.~{Porciani}, S.~J. {Lilly}, F.~{Miniati}, {\it \apj\/} {\bf 628}, 61 (2005).

\bibitem{Kimock2021}
B.~{Kimock}, {\it et~al.\/}, {\it \apj\/} {\bf 909}, 119 (2021).


\bibitem{Ferland2017}
G.~J. {Ferland}, {\it et~al.\/}, {\it Rev. Mex. Astro. Astrof.\/} {\bf 53}, 385 (2017).

\bibitem{Mathews1987}
W.~G. {Mathews}, G.~J. {Ferland}, {\it \apj\/} {\bf 323}, 456 (1987).

\bibitem{Thomas2016}
A.~D. {Thomas}, {\it et~al.\/}, {\it \apj\/} {\bf 833}, 266 (2016).

\bibitem{Shang2005}
Z.~{Shang}, {\it et~al.\/}, {\it \apj\/} {\bf 619}, 41 (2005).

\bibitem{Molina2013}
M.~{Molina}, {\it et~al.\/}, {\it \mnras\/} {\bf 433}, 1687 (2013).

\bibitem{Jin2012}
C.~{Jin}, M.~{Ward}, C.~{Done}, J.~{Gelbord}, {\it \mnras\/} {\bf 420}, 1825  (2012).


\bibitem{Natta1984}
A.~{Natta}, N.~{Panagia}, {\it \apj\/} {\bf 287}, 228 (1984).

\bibitem{Draine2011}
B.~T. {Draine}, {\it {Physics of the Interstellar and Intergalactic Medium}\/}, Princeton University Press (2011).

\bibitem{Salim2018}
S.~{Salim}, M.~{Boquien}, J.~C. {Lee}, {\it \apj\/} {\bf 859}, 11 (2018).

\bibitem{Prochaska2009}
J.~X. {Prochaska}, J.~F. {Hennawi}, {\it \apj\/} {\bf 690}, 1558 (2009).

\bibitem{Lau2016}
M.~W. {Lau}, J.~X. {Prochaska}, J.~F. {Hennawi}, {\it \apjs\/} {\bf 226}, 25  (2016).

\bibitem{Hennawi2015}
J.~F. {Hennawi}, J.~X. {Prochaska}, S.~{Cantalupo}, F.~{Arrigoni-Battaia}, {\it  Science\/} {\bf 348}, 779 (2015).

\bibitem{McCourt2018}
M.~{McCourt}, S.~P. {Oh}, R.~{O'Leary}, A.-M. {Madigan}, {\it \mnras\/} {\bf  473}, 5407 (2018).

\bibitem{Boquien2019}
M.~{Boquien}, {\it et~al.\/}, {\it \aap\/} {\bf 622}, A103 (2019).

\bibitem{Kennicutt1998}
J.~{Kennicutt}, Robert~C., {\it \araa\/} {\bf 36}, 189 (1998).

\bibitem{Lu2014}
Z.~{Lu}, {\it et~al.\/}, {\it \mnras\/} {\bf 439}, 1294 (2014).

\bibitem{Evrard2008}
A.~E. {Evrard}, {\it et~al.\/}, {\it \apj\/} {\bf 672}, 122 (2008).




\bibitem{Danovich2015}
M.~{Danovich}, A.~{Dekel}, O.~{Hahn}, D.~{Ceverino}, J.~{Primack}, {\it
  \mnras\/} {\bf 449}, 2087 (2015).
  
\bibitem{Teklu2015}
A.~F. {Teklu}, {\it et~al.\/}, {\it \apj\/} {\bf 812}, 29 (2015).

\bibitem{Wang2021}
S.~{Wang}, {\it et~al.\/}, {\it \mnras\/} {\bf 509}, 3148 (2021).


\bibitem{Martin2015}
D.~C. {Martin}, {\it et~al.\/}, {\it \nat\/} {\bf 524}, 192 (2015).

\bibitem{Ringermacher2009}
H.~I. {Ringermacher}, L.~R. {Mead}, {\it \mnras\/} {\bf 397}, 164 (2009).

\bibitem{Fox2019}
A.~J. {Fox}, {\it et~al.\/}, {\it \apj\/} {\bf 884}, 53 (2019).

\bibitem{Nesvadba2008}
N.~P.~H. {Nesvadba}, M.~D. {Lehnert}, C.~{De Breuck}, A.~M. {Gilbert}, W.~{van
  Breugel}, {\it \aap\/} {\bf 491}, 407 (2008).
  
\bibitem{Harrison2012}
C.~M. {Harrison}, {\it et~al.\/}, {\it \mnras\/} {\bf 426}, 1073 (2012).

\bibitem{Harrison2014}
C.~M. {Harrison}, D.~M. {Alexander}, J.~R. {Mullaney}, A.~M. {Swinbank}, {\it
  \mnras\/} {\bf 441}, 3306 (2014).

\bibitem{Cano-Diaz2012}
M.~{Cano-D{\'\i}az}, {\it et~al.\/}, {\it \aap\/} {\bf 537}, L8 (2012).

\bibitem{Greene2012}
J.~E. {Greene}, N.~L. {Zakamska}, P.~S. {Smith}, {\it \apj\/} {\bf 746}, 86  (2012).

\bibitem{Herrera-Camus2019}
R.~{Herrera-Camus}, {\it et~al.\/}, {\it \apj\/} {\bf 871}, 37 (2019).

\bibitem{Pequignot1991}
D.~{Pequignot}, P.~{Petitjean}, C.~{Boisson}, {\it \aap\/} {\bf 251}, 680  (1991).

\bibitem{Harrison2018}
C.~M. {Harrison}, {\it et~al.\/}, {\it Nature Astronomy\/} {\bf 2}, 198 (2018).

\bibitem{Popping2014}
G.~{Popping}, R.~S. {Somerville}, S.~C. {Trager}, {\it \mnras\/} {\bf 442},  2398 (2014).

\bibitem{Gultekin2021}
K.~{Gultekin}, {\it et~al.\/}, {\it \apj\/} {\bf 906}, 48 (2021).

\bibitem{Shang2011}
Z.~{Shang}, {\it et~al.\/}, {\it \apjs\/} {\bf 196}, 2 (2011).









  
  

  




  











  
  
  





















 
 

 
 




  

  

  




  






























\end{thebibliography}
\end{document}